\documentclass[useAMS,usenatbib,twocolumn]{mnras}

\usepackage{amsmath}
\usepackage{graphicx}
\usepackage{dcolumn}
\usepackage{bm}
\usepackage{epsfig}
\usepackage{amssymb,latexsym,mathrsfs}
\usepackage{graphicx}
\usepackage{color}
\usepackage{hyperref}

\usepackage{float}
\usepackage{natbib}

\usepackage[dvipsnames]{xcolor}

\hypersetup{
    colorlinks=true,
    linkcolor=red,
    citecolor=blue,
}

\def\apjl{{ ApJL}}
\def\apjs{{ ApJS}}

\def\aap{{ A\&A}}

\newcommand{\be}{\begin{equation}}
\newcommand{\ee}{\end{equation}}
\newcommand{\bs}{\begin{split}} 
\newcommand{\bea}{\begin{eqnarray}}
\newcommand{\eea}{\end{eqnarray}}

\title[Nucleosynthesis in PNS winds \& implications for UHECRs
]{On the synthesis of heavy nuclei in protomagnetar outflows and implications for ultra-high energy cosmic rays} 

\author[Bhattacharya et al.]{
Mukul Bhattacharya$^{1}$\thanks{mmb5946@psu.edu}, Shunsaku Horiuchi$^{2,3}$, Kohta Murase$^{1,4}$\\ 
${}^1$Department of Physics; Department of Astronomy \& Astrophysics; Center for Multimessenger Astrophysics,\\
Institute for Gravitation and the Cosmos, The Pennsylvania State University, University Park, PA 16802, USA\\
${}^2$Center for Neutrino Physics, Department of Physics, Virginia Tech, Blacksburg, VA 24061, USA\\
${}^3$Kavli IPMU (WPI), UTIAS, The University of Tokyo, Kashiwa, Chiba 277-8583, Japan \\
${}^4$Center for Gravitational Physics, Yukawa Institute for Theoretical Physics, Kyoto University, Kyoto, Kyoto 606-8502, Japan
}

\begin{document}

\date{Accepted . Received ; in original form }

\pagerange{\pageref{firstpage}--\pageref{lastpage}} \pubyear{2022}

\maketitle

\label{firstpage}

\begin{abstract} 
It has been suggested that 
strongly magnetised and rapidly rotating protoneutron stars (PNSs) may produce long duration gamma-ray bursts (GRBs) originating from stellar core collapse. 
We explore the steady-state properties and heavy element nucleosynthesis in neutrino-driven winds from such PNSs whose magnetic axis is generally misaligned with the axis of rotation. 
We consider a wide variety of central engine properties such as surface dipole field strength, initial rotation period and magnetic obliquity to show that heavy element nuclei can be synthesised in the radially expanding wind. This process is facilitated provided the outflow is Poynting-flux dominated such that its low entropy and fast expansion timescale enables heavy nuclei to form in a more efficient manner as compared to the equivalent thermal GRB outflows. 
We also examine the acceleration and survival of these heavy nuclei and show that they can reach sufficiently high energies $ \gtrsim 10^{20}\ {\rm eV}$ within the same physical regions that are also responsible for powering gamma-ray emission, primarily through magnetic dissipation processes. Although these magnetised outflows generally fail to achieve the production of elements heavier than lanthanides for our explored electron fraction range 0.4--0.6, we show that they are more than capable of synthesizing nuclei near and beyond iron peak elements.
\end{abstract}


\section{Introduction}
The origin and fate of intermediate and heavy nuclei\footnote{While intermediate mass nuclei have $A \sim 4-20$, heavy nuclei denotes elements that are heavier than $A \gtrsim 20$.} in ultrahigh-energy cosmic rays (UHECRs) is among one of the long standing unresolved problems in particle astrophysics (see \citealt{Blumer2009,KO2011,AB2019,Luis2019} for recent reviews). 
An extragalactic origin of the UHECR sources is supported by the observed flux cutoff at high energies $E \gtrsim 6\times10^{19}\, {\rm eV}$ measured by air shower observatories such as Pierre Auger Observatory (PAO; \citealt{PAO2015}), Telescope Array (TA; \citealt{AZ2013}) and High Resolution Fly's Eye (HiRes; \citealt{Abbasi2009}).
This is broadly consistent with the Greisen-Zatsepin-Kuzmin (GZK) cutoff or similar cutoff from the nuclear photodisintegration and Bethe-Heitler pair production processes due to the cosmic microwave background (CMB) and extragalactic background light (EBL) \citep{Abraham2008,Abbasi2008}. 
Furthermore, the observed anisotropic distributions for the arrival directions of UHECRs \citep{Abreu2010,Aab2015} suggest a possible correlation with nearby extragalactic gamma-ray sources such as starburst galaxies and AGN
\citep{Aab2018}, which further indicates that the UHECR sources trace the matter distribution within the GZK volume \citep{Aab2017}. 

The most commonly studied candidate sources of UHECRs comprise of persistent phenomena such as relativistic jets from active galactic nuclei (AGN; \citealt{Norman1995,Dermer2009,Peer2009,TH2011,Murase2017}), 
galaxy clusters (\citealt{Kang1996,Inoue2007,Murase2008,Kotera2009,FM2018})
and new-born pulsars \citep{Blasi2000,Arons2003,KM2009,Fang2014}, as well as transient phenomena that include classical gamma-ray bursts (GRBs; \citealt{Waxman1995,MU1995,KM2008a,Globus2015}), 
low-luminosity GRBs \citep{Murase2006,Murase2008,Zhang2018,Boncioli2019} 
and engine-driven supernovae \citep{Chakraborty2011,LW2012,ZM2019}, and tidal disruption events (TDEs; \citealt{FG2009,ABS2017,Zhang2017,Biehl2018,Gupin2018}). 
The diversity of these energetic events essentially originates from the characteristics of the central engine after incorporating stringent conditions on the magnetic field, size and energy that are required for accelerating the UHECRs. 

The composition of UHECRs provides important information regarding the origin of these energetic particles and is observationally inferred by estimating the shower depth at maximum elongation $X_{\rm max}$ as well as its RMS variation. 
Based on the recent measurements by PAO, the composition of UHECR nuclei is inferred to be increasingly dominated by heavier nuclei \citep{Abraham2010,Taylor2011,Abbasi2018,Batista2019} as pure proton or proton-helium compositions are generally disfavored at ultra-high energies \citep[e.g.,][]{Jiang2021,KT2021}.
If the composition of UHECRs is indeed dominated by an abundance of intermediate/heavy nuclei, this naturally provides various important implications towards constraining the UHECR sources. 
First, the source material itself should be sufficiently rich in intermediate/heavy nuclei, which suggests that the inner cores of compact/massive stars can be attractive. 
Second, the source environment must permit the intermediate/heavy nuclei to survive which may favor the magnetically-dominated models and/or low-luminosity jets with smaller bulk Lorentz factors over the typical high-luminosity GRB jets. The nuclei should also survive during their propagation from the source to Earth. Partial disintegration, both in the source and during propagation, will signify that the source composition is heavier than what is observed, thereby placing even more stringent conditions. Moreover, there should be adequate number of sources located in the vicinity of our Galaxy such that the observed UHECR energy flux can be explained without the UHECR nuclei getting photodisintegrated into free nucleons or daughter nuclei during propagating towards Earth. However, it should be noted that the source identification in the field-of-view
is more challenging due to the larger magnetic deflections experienced by the nuclei. 

In this work, we investigate the relativistic outflows generated from strongly magnetised and rapidly rotating PNS to be potential UHECR sources as their environments can contain large abundance of intermediate or heavy nuclei. 
The core collapse of a massive star leads to the formation of a hot PNS, which subsequently launches a shock wave. If this shock gets revived by neutrino heating and propagates across the stellar envelope, a successful supernova explosion occurs. 
As the PNS gradually cools down, it drives a neutrino-heated wind into the post-shock medium that recedes with the Kelvin-Helmholtz timescale $t_{\rm KH} \sim 10-100\ {\rm sec}$ \citep{BL1986,JM1996,QW1996}. The asymptotic heavy nuclei yields of these winds depend on the entropy and electron fraction of the wind matter, as well as dynamical expansion timescale of the expanding wind \citep{Pruet2006,Fisker2009,Wanajo2011}.

It has already been shown that nuclei can be accelerated to very high energies $E \gtrsim 10^{20}\ {\rm eV}$ and also survive in the jet dissipation regions for both high-luminosity GRBs and low-luminosity GRBs \citep{Murase2008,KM2010,Metzger2011b,Horiuchi2012}. 
Although the specific details of the energy dissipation mechanism are still debated, it is predicted using two main physical models: magnetic reconnection and internal shocks. While magnetic reconnections can occur if the outflow develops alternating field structure due to a general misalignment between the magnetic field and rotation axes, internal shocks can arise at large radial distances as the Lorentz factor of the wind gradually increases over time. If GRB jets are indeed powered by the magnetorotational energy of PNS, the products of nucleosynthesis in the wind will be present in the relativistic jet that escapes stellar ejecta and subsequently powers the prompt gamma-ray emission. 

Previous studies of neutrino-driven winds focused on non-rotating and spherically symmetric winds that are accelerated by thermal pressure could not achieve the necessary conditions for nucleosynthesis to reach the second and third rapid neutron capture (r-process) peaks \citep{Roberts2010,AM2011,Fischer2012,Roberts2012}. 
It is feasible for millisecond (ms) PNS winds to achieve favourable conditions for r-process nucleosynthesis to occur via strong magnetocentrifugal acceleration provided that the rapid rotation efficiently couples with the strong magnetic fields \citep[e.g.,][]{Metzger2008}. 
For outflows generated from strongly magnetised and rapidly rotating PNS, the total mass loss rate was found to be significantly lower compared to spherical winds as those outflows are confined to only a small fraction of PNS surface that is threaded by the open field lines \citep{Thompson2003,Metzger2007}. Recent studies have confirmed that relativistic outflows with a combination of high entropy, short dynamical expansion timescale and low electron fraction can potentially achieve nucleosynthesis up to the second and third r-process peaks with mass fractions of some elements enhanced by factors of $\sim 10^2-10^3$ \citep{Vlasov2014,Vlasov2017}. 
While the proposed mechanisms required in order to realise such heavy nuclei synthesis generally include postulating additional heating sources \citep{SN2005,Metzger2007} or considering extremely massive neutron stars \citep{Wanajo2013}, it is not yet clear as to how common or rare these necessary conditions might be.

In this work, we present a comprehensive study of the steady-state properties of neutrino-driven winds from strongly magnetised and rapidly rotating PNS. We explore the origin, survival and acceleration of UHECR nuclei in these magnetised outflows for a wide range of surface magnetic field, initial rotation period and magnetic obliquity. This paper is organised as follows. In Section~\ref{Sec2}, we briefly describe the evolutionary stages of the PNS wind and estimate the steady-state properties of the outflow for a specific set of initial parameters of the central engine. In Section~\ref{Sec3}, we discuss the physics of heavy element nucleosynthesis that takes place in the magnetised outflows originating from these systems. 
We then examine the necessary conditions for the r-process nucleosynthesis to potentially reach very heavy r-process nuclei in Section~\ref{Sec4}. Next, in Section~\ref{Sec5}, we consider the acceleration and survival of UHECR nuclei within the dissipation regions that are also responsible for powering gamma-ray emission via particle acceleration. Finally, we discuss the main implications of our results in Section~\ref{Sec6} and summarise our conclusions in Section~\ref{Sec7}.

\section{Evolution of Protomagnetar wind properties}
\label{Sec2}
In this section, we first briefly discuss the evolutionary stages of the neutrino-driven winds that originate from the surface of the rapidly rotating and strongly magnetised PNS. We then estimate the time-dependent properties of these magnetised outflows that is physically characterised by the long-term cooling evolution of the PNS \citep{Thompson2004,Metzger2007,Metzger2011a}.

\subsection{Stages of PNS evolution}
\label{Sec2.1}
Continued nuclear burning inside a star builds up a massive iron core, which leads to gravitational instability and resultant core collapse. A strong shock is then launched due to core bounce while the inner core forms a hot PNS that is primarily composed of neutrons. The vast majority of the gravitational binding energy is released in terms of neutrinos \citep{BL1986}.
The shock wave gradually loses energy due to the dissociation of iron nuclei and neutrino cooling close to PNS. In the ``neutrino mechanism'', the shock is revived due to neutrino heating in the gain region which lies behind the stalled shock wave (see, e.g. \citet{Couch2017,Mezz2005,Kotake2006,Janka2012,Burrows2013}, for the detailed mechanism). 
Shortly afterwards, a non-relativistic neutrino-heated wind is launched from the PNS surface. The wind then expands freely into the cavity that is already evacuated by the outgoing shock. 
Over the first few seconds, the neutrino luminosity is primarily driven by the cooling and contraction of the shock-heated outer layers of the PNS from an initial radius of $\sim 30\ {\rm km}$ to a final radius of $\sim 10\ {\rm km}$. The long-term cooling evolution is initiated once the PNS stops contracting and proceeds over $t \sim t_{\rm KH} \sim 30-100\ {\rm sec}$ \citep{QW1996,Pons1999}.

While most of the wind energy is stored as Poynting flux close to the PNS surface, the magnetic energy is gradually converted into the bulk kinetic energy of the wind, thereby powering its acceleration. As the PNS continues to cool, the magnetisation increases rapidly and the neutrino-driven winds achieve relativistic terminal bulk Lorentz factors. 
The interaction of the wind with the stellar ejecta leads to its collimation into a bipolar jet which then breaks out of the star after $t \sim t_{\rm bo}$. We estimate the jet-breakout time from the stellar ejecta based on the analytical fit given by \citet{Bromberg2015} 
\be 
t_{\rm bo} = 6.5\ {\rm s}\ \left[\left(\frac{\dot{E}_{\rm iso}}{\dot{E}_{\rm rel}}\right)^{-2/3} + \left(\frac{\dot{E}_{\rm iso}}{\dot{E}_{\rm rel}}\right)^{-2/5}\right]^{1/2}
\ee
where $\dot{E}_{\rm iso}$ is the isotropic jet luminosity and $\dot{E}_{\rm rel} \approx 1.6\times10^{49}\ {\rm erg/s}$ is the transition luminosity between the non-relativistic and relativistic breakout times for typical stellar structure parameters. 
This fit for $t_{\rm bo}$ assumes a Poynting-flux dominated jet with an opening angle $\theta_j \approx 7^{\circ}$ from a $15\, M_{\odot}$ and $4\, R_{\odot}$ Wolf-Rayet star with stellar density profile $\rho \propto r^{-2.5}$.
For the range of PNS initial parameters that we consider here, this further yields $t_{\rm bo}= 4.2\ {\rm s}\ (B_{\rm dip}/3\times10^{15}\ {\rm G})^{-1/3}(P_0/3\ {\rm ms})$. 

Subsequently, the relativistic wind propagates through a relatively sparse region outside the star and the GRB phase is initiated once the outflow reaches the dissipation radius. 
The GRB emission can be powered by two distinct mechanisms: (a) the dissipation of jet's Poynting flux near the photosphere due to magnetic instabilities or misalignment, (b) the internal shocks that occur within the jet at larger radii owing to the increasing wind Lorentz factor. 
Once the PNS becomes optically thin to neutrinos, which typically occurs after $t_{\nu,\rm thin} \sim 30-100\ {\rm sec}$, the wind magnetization significantly increases and thereby particle acceleration via dissipation processes become inefficient. Thus, the wind can no longer power the prompt GRB emission.
It should be noted that the neutrino cooling phase and transparency timescales are both sensitive to the properties of neutrino opacities within the dense PNS matter \citep{BS1998,Hudepohl2010,Roberts2012}. 
The precise value of $t_{\nu,\rm thin}$ depends on the unknown NS equation of state as well as the rotation rate ($t_{\nu,\rm thin} \propto \eta_s$) and weak-physics. 
As the exact dependence of neutrino opacity on the PNS wind properties and neutrino cooling curves (see Figure~\ref{Fig_neuLE}) is fairly uncertain, here we assume a fixed $t_{\nu,\rm thin} \approx 77\, {\rm sec}$ when the neutrino luminosity drops to 1\% of its initial value (see also \citet{Metzger2011a}).

\subsection{Neutrino cooling curves}
\label{Sec2.2}
As the mass loss from the PNS surface is driven by neutrino heating, it depends sensitively on the neutrino luminosity $L_{\nu}$ and the mean neutrino energy $\epsilon_{\nu}$ during the Kelvin-Helmholtz cooling phase \citep{QW1996} 
\bea
\dot{M} = 5\times10^{-5} M_{\odot}{\rm s^{-1}}f_{\rm open} f_{\rm cent} \left(\frac{L_{\nu}}{10^{52}\ {\rm erg/s}}\right)^{5/3}  \nonumber
\\ \times \left(\frac{\epsilon_{\nu}}{10\ {\rm MeV}}\right)^{10/3} \left(\frac{M_{\rm ns}}{1.4 M_{\odot}}\right)^{-2} \left(\frac{R_{\rm ns}}{10\ {\rm km}}\right)^{5/3} (1+\epsilon_{es})^{5/3}
\label{Mdot}
\eea
where $M_{\rm ns}$ is the mass, $R_{\rm ns}$ is the radius, $f_{\rm open}$ is the fraction of PNS surface that is threaded by open magnetic field lines and $f_{\rm cent}$ accounts for the  magnetocentrifugal slinging effect. To study the variation of $L_{\nu}$, $\epsilon_{\nu}$ and $R_{\rm ns}$ over time, we will use the results from \citet{Pons1999} cooling models that are evaluated for $M_{\rm ns}=1.4\ M_{\odot}$ (see Appendix \ref{AppendixA} for discussion). The parameter $\epsilon_{es} = 0.2(M_{\rm ns}/1.4\ M_{\odot})(R_{\rm ns}/10\ {\rm km})^{-1}(\epsilon_{\nu}/10\ {\rm MeV})^{-1} \lesssim 1$ incorporates correction for the additional heating that occurs due to inelastic electron scattering events. 

It should be noted that \citet{Pons1999} did not consider the effects of rotation and convection on the PNS cooling evolution and the effects of strong magnetic fields on the neutrino-driven mass loss rate. 
While strong fields will suppress the mass loss rate by a factor $f_{\rm open}$ as only the open fraction of the surface contributes to the outflow, $\dot{M}$ also gets enhanced by a factor $f_{\rm cent}$ -- particularly for rapid rotation rate and large magnetic obliquity $\chi$,  owing to the increased scale height within the heating region \citep{Thompson2004}. 
Significantly rapid rotation decreases the PNS interior temperature which then slows down its cooling evolution. 
To qualitatively account for the effects of rotation, here we include a stretch factor $\eta_s \sim 3$ that appropriately modifies the cooling evolution relative to the non-rotating case as \citep{Thompson2005,Metzger2011a}
\be
L_{\nu} \rightarrow{L_{\nu}|_{\Omega=0} \eta_s^{-1}};\ \ t \rightarrow{t|_{\Omega=0} \eta_s};\ \ \epsilon_{\nu} \rightarrow{\epsilon_{\nu}|_{\Omega=0} \eta_s^{-1/4}}
\ee
where $t$ is the time since core-collapse.
Although we consider a fixed value for $\eta_s$ here, ideally it should be an increasing function of the PNS rotation rate. 

The maximum centrifugal enhancement to the mass loss rate is approximated by $f_{\rm cent,max}={\rm exp}[(P_{\rm cent}/P)^{1.5}]$ where $P_{\rm cent} = (2.1\ {\rm ms})\ {\rm sin}\ \alpha \ (R_{\rm ns}/10\ {\rm km})^{3/2} (M_{\rm ns}/1.4 M_{\odot})^{-1/2}$, $\alpha = {\rm max}(\theta_{\rm open}/2,\chi)$, $\theta_{\rm open}$ is the opening angle of the polar cap and $\chi$ is the inclination angle between the magnetic field and rotational axes (see \citealt{Metzger2008}). Based on the numerical results of \citet{Metzger2008}, we assume that the mass loss enhancement factor is $f_{\rm cent} = f_{\rm cent,max}(1 - {\rm exp}[-R_A/R_s]) + {\rm exp}[-R_A/R_s]$ where $R_A = R_L{\rm min}(\sigma_0^{1/3},1)$ is the Alfven radius, $R_L = c/\Omega$ is the light cylinder radius and $R_s = (GM/\Omega^2)^{1/3}$ is the sonic radius \citep{Metzger2007,LC1999}.

\begin{figure} 
\includegraphics[width=1.03\columnwidth]{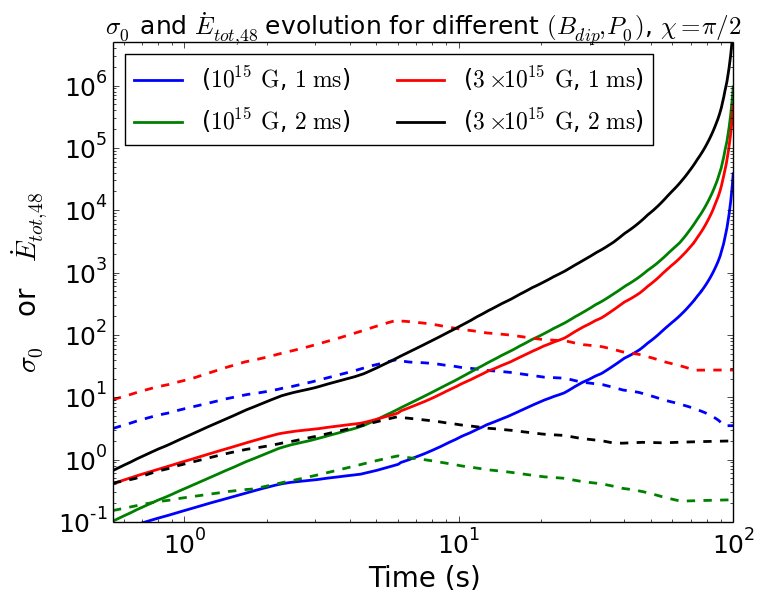} 
\vspace{-0.5cm}
\caption{The effect of surface dipole magnetic field $B_{\rm dip}$ and initial rotation period $P_0$ on the time evolution of 
PNS wind magnetisation $\sigma_0$ (solid) and energy loss rate $\dot{E}_{\rm tot}=\dot{E}_{\rm tot,48}\times10^{48}\,{\rm erg/s}$ (dashed), all shown for magnetic obliquity angle $\chi=\pi/2$. In general, larger $B_{\rm dip}$ increases both $\sigma_0$ and $\dot{E}_{\rm tot}$, while larger $P_0$ increases $\sigma_0$ but decreases $\dot{E}_{\rm tot}$.} 
\label{sigma0_Edot}
\end{figure}

\subsection{Neutrino-driven wind}
\label{Sec2.3}
The rotational energy of the PNS can be written as
\begin{equation}
E_{\rm rot} = \frac{1}{2}I\Omega^2 = 3\times10^{52}\ {\rm ergs}\left(\frac{M_{\rm ns}}{1.4M_{\odot}}\right)\left(\frac{R_{\rm ns}}{12\ {\rm km}}\right)^{2} \left(\frac{P}{{\rm ms}}\right)^{-2}
\end{equation}
where $I=(2/5)M_{\rm ns}R_{\rm ns}^{2}$ is the moment of inertia and $\Omega=2\pi/P$ is the PNS rotation rate corresponding to a period $P$. 
The variation in the wind properties are largely driven by the increase in magnetisation $\sigma_0 = \phi_B^2 \Omega^2/\dot{M}c^3$ over time. 
The magnetic flux due to a rotating dipole with surface magnetic field $B_{\rm dip}$ is given by $\phi_B = (f_{\rm open}/4\pi)B_{\rm dip}R_{\rm ns}^2$.  
The total wind power in the asymptotic limit can be written as a combination of just the kinetic energy and magnetic flux components $\dot{E}_{\rm tot} = \dot{E}_{\rm kin} + \dot{E}_{\rm mag}$. 
The wind magnetisation $\sigma_0$ is critical as it affects the asymptotic partition between the kinetic and magnetic energies in the wind. 
The wind kinetic luminosity is $\dot{E}_{\rm kin} = (\Gamma_{\infty}-1)\dot{M}c^2$, where $\Gamma_{\infty}$ is determined by the asymptotic outflow velocity. 
While non-relativistic outflows ($\sigma_0 \lesssim 1$) reach asymptotic speed $v_{\infty}\approx c\sigma_0^{1/3}$ resulting in $\dot{E}_{\rm kin} = (1/2)\dot{M}v_{\infty}^2 \propto \dot{M}^{1/3}$, relativistic outflows ($\sigma_0 \gtrsim 1$) achieve $\Gamma_{\infty} \approx \sigma_0^{1/3}$ whereby $\dot{E}_{\rm kin}$ turns out to be independent of $\dot{M}$. Magnetic power $\dot{E}_{\rm mag}$ is directly related to the flux $\phi_B$ and therefore to the magnetisation $\sigma_0$ as
\be
\dot{E}_{\rm mag} \approx \frac{2}{3} \left\{
\begin{array}{ll}
\dot{M}c^2 \sigma_0^{2/3}, & \sigma_0 \ll 1 \vspace{0.2cm} \\ 
\dot{M}c^2 \sigma_0, & \sigma_0 \gg 1 \\
\end{array}
\right.
\label{E_mag}
\ee
The kinetic and magnetic contributions to the total power are similar $\dot{E}_{\rm mag}/\dot{E}_{\rm kin} =2$ in the non-relativistic case, whereas most of the wind power resides in Poynting flux $\dot{E}_{\rm mag}/\dot{E}_{\rm kin} \sim \sigma_0^{2/3} \gg 1$ for relativistic outflows. The magnitude of $\sigma_0$ decides the efficiency with which the outflow can accelerate and dissipate its energy, and equals the maximum achievable Lorentz factor $\Gamma_{\rm max} \approx \dot{E}/\dot{M}c^2 \approx \sigma_0$.

Figure \ref{sigma0_Edot} shows the effect of surface magnetic field $B_{\rm dip}$ and initial rotation period $P_0$ on the magnetisation $\sigma_0$ and energy loss rate $\dot{E}_{\rm tot}$, calculated for four wind models with $(B_{\rm dip},P_0,\chi)= (10^{15}\, {\rm G},1\, {\rm ms},\pi/2),\, (10^{15}\, {\rm G},2\, {\rm ms},\pi/2),\, (3\times10^{15}\, {\rm G},1\, {\rm ms},\pi/2)\, {\rm and}\, (3\times10^{15}\, {\rm G},2\, {\rm ms},\pi/2)$. The time evolution of $\sigma_0$ is qualitatively similar for all cases as the initially non-relativistic ($\sigma_0 < 1$) wind corresponding to a high neutrino-driven mass loss rate gradually evolves to a relativistic ($\sigma_0 \gtrsim 1$) outflow over the first few seconds. We find that stronger fields and slower PNS rotation rates result in a larger jet magnetisation. While the rotation period $P_0$ has a noticeable effect on the slope of the $\sigma_0 (t)$ curve, the dipole field $B_{\rm dip}$ only changes the relative normalisation. For each wind model, $\dot{E}_{\rm tot}$ increases for the first few seconds as $B_{\rm dip}$ and $\Omega$ increase due to magnetic flux and angular momentum conservation, respectively, over the course of PNS radial contraction. The wind power attains its maximum value around $t \sim 5-10\, {\rm sec}$ and then decreases as the PNS spins down with a gradually shrinking magnetosphere. We find that $\dot{E}_{\rm tot}$ rises with larger $B_{\rm dip}$ and smaller $P_0$, i.e., with stronger magnetocentrifugal acceleration.

\begin{figure*} 
\includegraphics[width=0.49\textwidth]{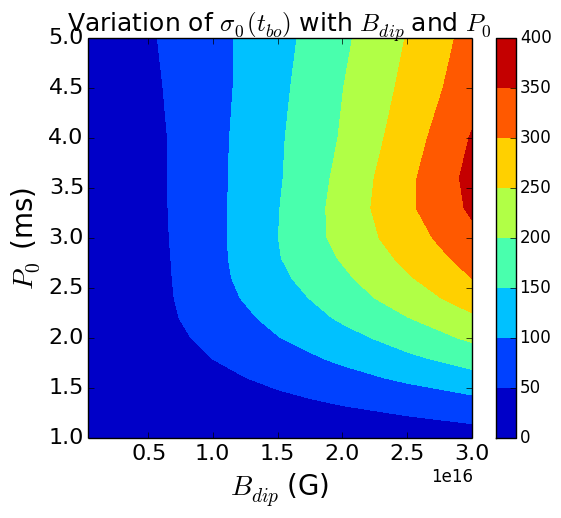} 
\includegraphics[width=0.49\textwidth]{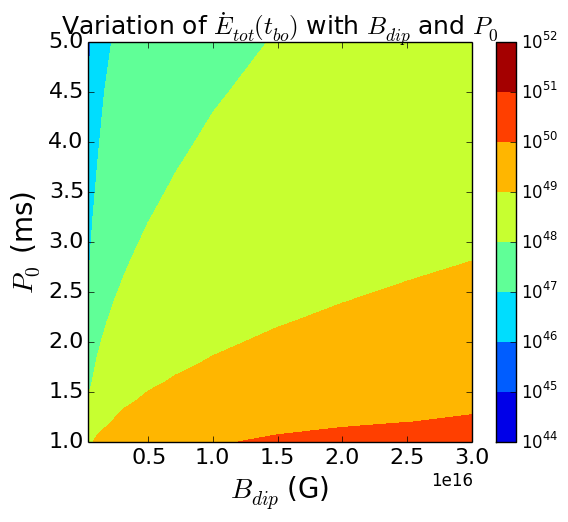}
\vspace{-0.2cm}
\caption{Contour plots for the wind magnetisation $\sigma_0$ (left panel) and energy loss rate $\dot{E}_{\rm tot}$ (right panel), shown as functions of the surface dipole magnetic field $B_{\rm dip}$ and initial rotation period $P_0$, both at the time of jet breakout $t_{\rm bo}$. For the $B_{\rm dip}$ and $P_0$ ranges shown, $t_{\rm bo}$ falls within $\sim 6-15\, {\rm s}$. As in Figure~\ref{sigma0_Edot}, we have assumed fixed magnetic obliquity angle $\chi=\pi/2$.}
\label{Ctr_sigma0_Edot}
\end{figure*}

The development of hydrodynamic instabilities in weakly magnetised jets can lead to jet-cocoon mixing which increases the baryon loading in the jet and reduces its Lorentz factor at early times. A sub-dominant magnetisation $10^{-2} < \sigma_0 < 1$ can stabilise the jet against these hydrodynamic instabilities on the jet-cocoon boundary and reduce the baryonic mixing with the cocoon \citep{Gottlieb2021a}. Numerical studies of Poynting-flux dominated jets (see \citealt{BT2016,Matsumoto2021}) have also shown that magnetic kink instability can reduce the magnetisation to $\sigma_0 \sim 1$, although modelling such highly magnetized jets typically requires a higher resolution compared to hydrodynamic jets \citep{Gottlieb2021b}. Here, we do not consider the effect of jet-cocoon mixing on $\sigma_0$ as it is not expected to affect the abundance of intermediate or heavy nuclei from r-process nucleosynthesis that occurs on longer timescales $t \sim  t_{\rm KH} \gtrsim t_{\rm bo}$. After the breakout, the jet could be more strongly magnetized and more relativistic, which may also affect high-energy neutrino production in jets inside stars.

It should be noted that magnetic heating close to the protomagnetar surface can also lead to uncertainties in baryon loading and thereby to the magnetisation $\sigma_0$. Outflows that are driven by viscous heating from the accretion disk can also become efficient once the neutrino cooling slows down (see recent numerical relativity simulation results from \citealt{Shibata21}). In addition, neutron diffusion can also turn out to be more efficient for highly magnetized jets \citep{Preau2021}. However, detailed studies of these effects is beyond the scope of this work.

\subsection{Spin-down evolution}
\label{Sec2.4}
A rotating neutron star gradually loses its angular momentum $J=(2/5)M_{\rm ns}R_{\rm ns}^2 \Omega$ to the wind at the rate $\dot{J} = -\dot{E}/\Omega$. The time evolution of the angular velocity $\Omega$ is then obtained directly from $\dot{\Omega}/\Omega = -2(\dot{R}_{\rm ns}/R_{\rm ns}) -0.5(\dot{E}/E_{\rm rot}$). 
The solution for the PNS spin-down evolution is completely specified by its mass $M_{\rm ns}$, surface dipole field $B_{\rm dip}$, initial rotation period $P_0 = 2\pi/\Omega_0$ and magnetic obliquity $\chi$. 
As the PNS continues to contract for a few seconds post core bounce, we define $\Omega_0$ and $B_{\rm dip}$ as the maximum values that are achieved provided the PNS contracts with a conserved angular momentum $J \propto R_{\rm ns}^2 M_{\rm ns} \Omega$ and magnetic flux $\phi \propto B_{\rm dip}R_{\rm ns}^2$.

Here we consider $M_{\rm ns}=1.4\ M_{\odot}$, $3\times10^{14}\ {\rm G} \lesssim B_{\rm dip} \lesssim 3\times10^{16}\ {\rm G}$, $1\ {\rm ms} \lesssim P_0 \lesssim 5\ {\rm ms}$ and $\chi=\pi/2$. The effect of varying obliquity $\chi$ and NS equation of state (and thereby $M_{\rm ns}$) are discussed in Appendix \ref{AppendixA}. Although $B \sim 3\times10^{17}\ {\rm G}$ is possible with $\epsilon_B \sim 1$ (fraction of the rotational energy contained in Poynting flux), stable field configurations require the dipole field component $B_{\rm dip}$ to be $<10\%$ of the total field strength. Furthermore, significantly large fields $B_{\rm dip} \gtrsim 3\times10^{16}\ {\rm G}$ can potentially affect the neutrino-driven mass loss rate obtained from equation (\ref{Mdot}). 
The left and right panels of Figure \ref{Ctr_sigma0_Edot} show the contours for $\sigma_0$ and $\dot{E}_{\rm tot}$ respectively, computed at the jet breakout time $t_{\rm bo}$, as functions of the surface dipole field $B_{\rm dip}$ and initial rotation period $P_0$. We find that $\sigma_0$ has a weak dependence on the rotation rate except for rapidly spinning cases with $P_0 \lesssim 2\, {\rm ms}$, for which $\sigma_0$ is suppressed significantly due to the enhanced mass loss aided by magnetocentrifugal slinging effect. With an increase in $B_{\rm dip}$, the magnetisation scales up as $\sigma_0 \propto \phi_B^2$. While weakly magnetised outflows with $B_{\rm dip} \lesssim 3\times10^{15}\, {\rm G}$ are non-relativistic at breakout, outflows with $B_{\rm dip} \gtrsim 1.5\times10^{16}\, {\rm G}$ and $P_0 \gtrsim 2.5\, {\rm ms}$ result in highly relativistic jets. The wind power $\dot{E}_{\rm tot}$ is dominated by the magnetic power $\dot{E}_{\rm mag}$ for outflows with large magnetisation. We find that stronger magnetic fields and faster rotation rates both assist in elevating $\dot{E}_{\rm tot}$ considerably, sometimes even by many orders of magnitude. As a result, the feasibility of successful jets turns out to be much higher in outflows with $B_{\rm dip} \gtrsim 10^{16}\, {\rm G}$ and $P_0 \lesssim 2\, {\rm ms}$.

Late-time accretion onto the PNS can affect the spindown evolution by altering the magnetosphere geometry and through accretion torques \citep{Metzger2008,ZD2009}. Angular momentum accretion from the disk assists the PNS to remain rapidly spinning which effectively enhances the mass loss rate and also reduces the wind magnetisation \citep{Thompson2004,Metzger2007}. We discuss the consequence of angular momentum accretion through an equatorial disk as well as dipolar magnetic field growth aided by differential rotation on the long-term cooling evolution of the PNS in Appendix \ref{AppendixB}.

\begin{figure} 
\includegraphics[width=\columnwidth]{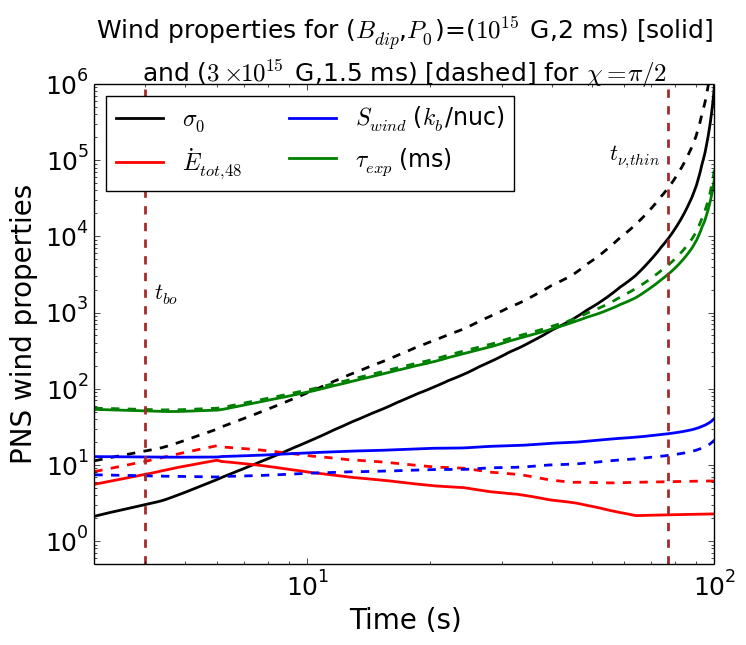} 
\vspace{-0.5cm}
\caption{The time evolution of wind entropy $S_{\rm wind}$ and expansion timescale $t_{\rm exp}$, as well as the outflow magnetisation $\sigma_0$ and energy loss rate $\dot{E}_{\rm tot}$ are shown. The wind properties are evaluated for PNS systems with $(B_{\rm dip},P_0,\chi)=(10^{15}\, {\rm G},2\, {\rm ms},\pi/2)$ [solid curves] and $(3\times10^{15}\, {\rm G},1.5\, {\rm ms},\pi/2)$ [dashed curves]. The breakout time shown here is for $(B_{\rm dip},P_0)=(10^{15}\, {\rm G},2\, {\rm ms})$; it is somewhat shorter for $(B_{\rm dip},P_0)=(3 \times 10^{15}\, {\rm G},1.5\, {\rm ms})$.
} 
\label{Swind_texp}
\end{figure}

\section{Heavy element nucleosynthesis}
\label{Sec3}
Neutrino-driven winds from rapidly rotating and highly magnetised PNS have been investigated as potential sites for intermediate and heavy element nucleosynthesis via the r-process \citep{Metzger2008,Metzger2011b}. 
While high-luminosity GRB jets may contain nuclei, Poynting flux dominated outflows or low-luminosity jets with smaller Lorentz factors provide more favourable conditions to synthesise significant heavy nuclei. 
Considering the magnetised jets that arise from a PNS central engine, \citet{Metzger2011b} showed that the wind composition may indeed be dominated by heavy nuclei with masses $A \sim 40-120$ that are formed as the hot material gradually expands outwards. The synthesis of heavier nuclei from the r-process requires an outflow with a combination of low entropy and short jet expansion timescale which is facilitated by the strong magnetocentrifugal acceleration generally found in rapidly rotating magnetars \citep{Metzger2008}. Similar considerations will also apply to relativistic winds that are powered by accretion, provided that the jet is magnetically-dominated rather than a thermally-driven fireball.

\subsection{Wind entropy and expansion timescale}
\label{Sec3.1}
For Poynting flux dominated outflows from the PNS, most of the energy at smaller radii is stored inside magnetic fields and the outflow has a significantly lower entropy $S_{\rm wind} \gtrsim 10-300\, {\rm k_b\, nucleon^{-1}}$ as compared to $S_{\rm wind} \gtrsim 10^5\, {\rm k_b\, nucleon^{-1}}$ for the thermally-driven fireball \citep{Eichler1989}. 
The equilibrium temperature near the PNS surface is set by equating the neutrino heating and cooling rates, and is generally $T \sim 1-2\, {\rm MeV}$ for the Kelvin-Helmholtz timescale $t \sim 10-100\, {\rm s}$. Gradually the hot jet material expands from the PNS surface and heavier elements start to form as lower temperatures are achieved at larger radii in the outflow.

As $T \propto 1/r$, the free nuclei recombine into Helium ($T = T_{\rm rec} \sim 0.5-1\, {\rm MeV}$) once the deuterium bottleneck is broken, which occurs a few $R_{\rm ns}$ above the PNS surface. 
Because this happens at low jet densities for high entropy where the further processing into carbon and heavier nuclei are not rapid enough compared to the expansion timescale, only a few elements heavier than He are formed, similar to Big Bang nucleosynthesis \citep{Lemoine2002,Pruet2002,Beloborodov2003}. However, the situation is considerably different for magnetically dominated jets from the PNS as the entropy is lower and He recombination occurs at higher densities, such that heavier nuclei can form efficiently via the triple-$\alpha$ process and subsequent $\alpha$ captures. 

Since the heavy elements form only after recombination, the abundance distribution of the synthesised nuclei is determined by the entropy $S_{\rm wind}$, expansion timescale $t_{\rm exp}$ and electron fraction $Y_e$ of the outflow at the recombination radius. The wind entropy is obtained from the neutrino heating that occurs in the gain region above the PNS surface but below the recombination radius and is approximated for slowly rotating and/or weakly magnetised PNS by (see \citealt{QW1996})
\bea
S_{\rm wind}(\Omega=0) = (88.5\, {\rm k_b/ nucleon})\ C_{es}^{-1/6}\left(\frac{L_{\nu}}{10^{52}\, {\rm erg/s}}\right)^{-1/6}\nonumber \\
\times \left(\frac{\epsilon_{\nu}}{10\, {\rm MeV}}\right)^{-1/3}
\left(\frac{M_{\rm ns}}{1.4\, M_{\odot}}\right)
\left(\frac{R_{\rm ns}}{10\, {\rm km}}\right)^{-2/3}
\eea
where $C_{es} = 1 + 0.7(M_{\rm ns}/1.4\, M_{\odot})(R_{\rm ns}/10\, {\rm km})^{-1}$ is a correction to the heating rate due to inelastic electron scattering. We have assumed that the electron neutrinos and antineutrinos have similar luminosities and mean energies, and included a $20\%$ enhancement in the entropy due to general relativistic gravity \citep{CF1997}. 

The wind entropy also depends on the obliquity angle $\chi$ as for the aligned rotators ($\chi=0$), the material is ejected near the rotational pole and the entropy is similar to the non-rotating value. The oblique rotators ($\chi=\pi/2)$, however, lose most of the matter from the rotational equator which reduces the heating experienced by the outflow and suppresses the entropy exponentially $S_{\rm wind}(\Omega,\chi=\pi/2) = S_{\rm wind}(\Omega=0)\,{\rm exp}(-P_{\rm cent}/P)$. The wind expansion timescale is $t_{\rm exp}=r/v_r$ evaluated at the recombination temperature $T=T_{\rm rec}$, where $v_r$ is the outflow velocity and $T_{\rm rec}$ is calculated from the outflow density $\rho$ and entropy $S_{\rm wind}$ assuming nuclear statistical equilibrium. The outflow velocity $v_r$ is obtained from the mass continuity relation $\dot{M}=\rho v_r A$, where $A = 4\pi R_{\rm ns}^2 f_{\rm open}(r/R_{\rm ns})^3$ is the area of the dipolar flux tube \citep{Metzger2011b}.

Figure \ref{Swind_texp} shows the time evolution of entropy $S_{\rm wind}$ and expansion timescale $t_{\rm exp}$ at the recombination radius, and their comparison with $\sigma_0$ and $\dot{E}_{\rm tot}$ of the outflow for $(B_{\rm dip},P_0,\chi)=(10^{15}\, {\rm G},2\, {\rm ms},\pi/2)\, {\rm and}\, (3\times10^{15}\, {\rm G},1.5\, {\rm ms},\pi/2)$. The jet breakout and the neutrino transparency timescales are also shown for reference. We find that $S_{\rm wind}$ stays roughly constant over the PNS cooling phase as the increase in value of $S(\Omega=0)$ is compensated by the exponential suppression factor. This is because equatorial outflows with $\chi=\pi/2$ experience significant centrifugal acceleration which reduces the heating received by the outflowing material and therefore its entropy. Although the expansion timescale increases gradually for $t \lesssim {\rm few}\, 10$s, at late times $t \sim t_{\nu,\rm thin}$, $t_{\rm exp}$ becomes comparable to the timescale over which the wind magnetisation changes. As expected, the increased magnetocentrifugal acceleration from outflows with stronger fields and rapid rotation rates result in larger $\sigma_0$ and larger $\dot{E}_{\rm tot}$ but correspondingly smaller $S_{\rm wind}$ due to the higher suppression factor. However, $t_{\rm exp}$ at the recombination radius is relatively unaffected by the outflow properties except for very late times $t \sim t_{\nu,\rm thin}$.

\begin{figure} 
\includegraphics[width=\columnwidth]{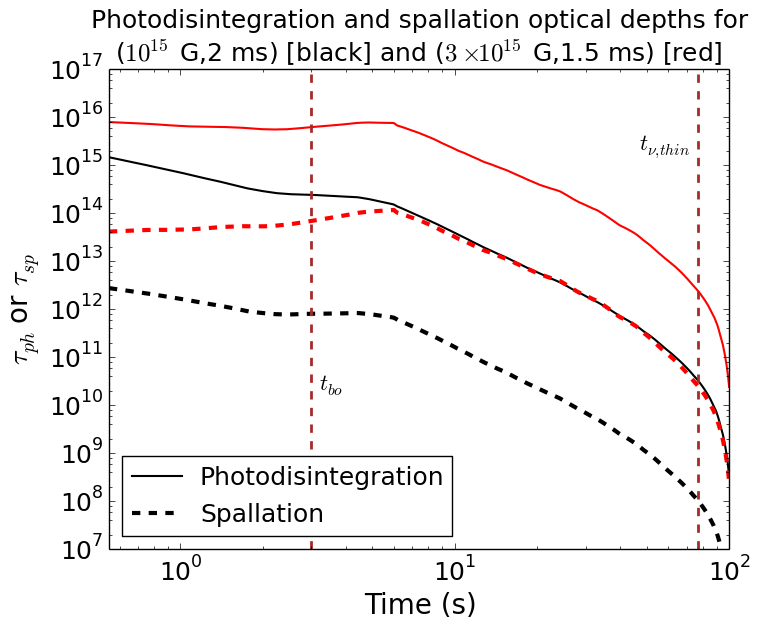} 
\vspace{-0.5cm}
\caption{The optical depths for photodisintegration (solid) and spallation (dashed) of nuclei injected at initial loading, which we take to occur just outside the PNS radius. The PNS radius shrinks from tens of km to about $12$ km over the time scales shown. 
The results here are obtained for PNS with $(B_{\rm dip},P_0,\chi)=(10^{15}\, {\rm G},2\, {\rm ms},\pi/2)$ [black curves] and $(3\times10^{15}\, {\rm G},1.5\, {\rm ms},\pi/2)$ [red curves]. 
The breakout time shown here corresponds to the $(B_{\rm dip},P_0)=(10^{15}\, {\rm G},2\, {\rm ms})$ model.
} 
\label{tau_photo_spal}
\end{figure}

\subsection{Initial composition}
\label{Sec3.2}
Nucleosynthesis in the wind takes place over $t \sim t_{\rm KH}$ only if baryons loaded onto the jet are disintegrated into free nucleons. Here we discuss the destruction of the nuclei through photodisintegration and spallation processes just above the PNS surface ($r/\Gamma \sim R_{\rm ns} \approx 12\, {\rm km}$) prior to the onset of recombination at larger outflow radii. Depending on the PNS outflow parameters, the nuclei can either survive these processes or get disintegrated into free nucleons. The typically high temperatures $T \gtrsim 1\, {\rm MeV}$ near the central engine indicates that most of the nuclei indeed get dissociated into free neutrons and protons.

The optical depth for photodisintegration is $\tau_{A\gamma} = n_{\gamma}\sigma_{A\gamma}r/\Gamma$, where $n_{\gamma}$ is the photon number density, $\sigma_{A\gamma} \approx 8\times10^{-26}\, {\rm cm^2}$ is the photodisintegration cross section and $r/\Gamma$ is the characteristic outflow radius in the comoving frame. 
The photon temperature is directly related to the radiation energy density as $aT^4 \approx L_{\gamma}/4\pi r^2 \Gamma^2 c \sim \epsilon_r \dot{E}_{\rm iso}/4\pi R_{\rm ns}^2 c$, where $\epsilon_r \sim 0.5$ is the radiative efficiency of the jet (see \citealt{Metzger2011a}). For our calculation here, we assume that the wind is collimated into a jet with opening angle $\theta_j \sim 4^{\circ}$ \citep{Bucciantini2009} such that the isotropic jet luminosity $\dot{E}_{\rm iso}$ is larger than the wind power $\dot{E}_{\rm tot}$ by a factor $f_b^{-1}$, where $f_b = \theta_j^2/2 \approx 2\times10^{-3}$ is the beaming fraction \citep{Rhoads1999}. 
Considering a thermal photon spectrum, the number density of photons is given by $n_{\gamma} \approx 19.23(kT/hc)^3$. 

Similarly, the optical depth for spallation can be written as $\tau_{sp} = n_0 \sigma_{sp} r/\Gamma$ where $n_0 = \dot{E}/4\pi R_{\rm ns}^2 \sigma_0 \overline{A} m_p c^3$ is the comoving ion density and $\overline{A} \sim 1$ is the average mass number of the baryons in the jet. The spallation cross section is $\sigma_{sp} = \sigma_0 A^{2/3}$ with $\sigma_0 \approx 3\times10^{-26}\, {\rm cm^2}$. We require both $\tau_{A\gamma} < 1$ and $\tau_{sp} < 1$ for the survival of nuclei during initial loading. In Figure \ref{tau_photo_spal}, the variation of the photodisintegration and spallation optical depths over time are shown for two different wind models $(B_{\rm dip},P_0,\chi)=(10^{15}\, {\rm G},2\, {\rm ms},\pi/2)\, {\rm and}\, (3\times10^{15}\, {\rm G},1.5\, {\rm ms},\pi/2)$. 
It should be noted that we evaluate the interaction optical depths at the radius where the nuclei are initially loaded onto the jet. The nuclei loading onto the jet typically occurs close to the PNS surface.
For spallation, we adopt $\overline{A} \sim 1$ as it gives the largest possible target density. For both the wind models considered here, we find that photodisintegration is significantly more important than spallation ($\tau_{A\gamma}/\tau_{sp} \sim 10^2-10^3$) as the larger number density of photons compared to ions compensates for the slightly smaller photodisintegration cross section. Furthermore, as $\tau_{A\gamma},\, \tau_{sp} \gg 1$, all the nuclei get rapidly disintegrated into free nucleons just above the PNS surface and can later participate in recombination at larger radii. We find that both $\tau_{A\gamma}$ and $\tau_{sp}$ are larger for stronger field and faster rotation, which is expected as the photon and ion number densities increase with the wind power.

\subsection{In-situ nucleosynthesis}
\label{Sec3.3}
In non-rotating and non-magnetised PNS winds, the electron fraction $Y_e$ at equilibrium is determined by the relative luminosities and mean energies of the electron-type neutrinos that diffuse out of the PNS interior \citep{QW1996}. 
If rotation and magnetic fields are dynamically important during the cooling phase, the magnetocentrifugal acceleration of the matter away from the PNS can lower $Y_e$ of the unbound ejecta by preventing the outflowing matter from coming into equilibrium with neutrino absorption reactions \citep{MQT08}. 
Although the temporal evolution of $Y_e$ depends on theoretical uncertainties in the neutrino spectrum, recent PNS cooling models (see \citealt{Hudepohl2010,R2012,Mirizzi2016}) indicate that the wind may attain moderately high $Y_e \approx 0.4-0.6$ while propagating outwards once the nucleons are irradiated with neutrinos whereas they are considerably more proton-rich at early times of the cooling epoch. Here we neglect the effects of strong magnetic fields on the neutrino heating and cooling rates in the PNS atmosphere as the corrections due to Landau quantization are found to be small for $B_{\rm dip} \lesssim 3\times10^{16}\, {\rm G}$ \citep{DQ2004}.

\begin{figure} 
\includegraphics[width=\columnwidth]{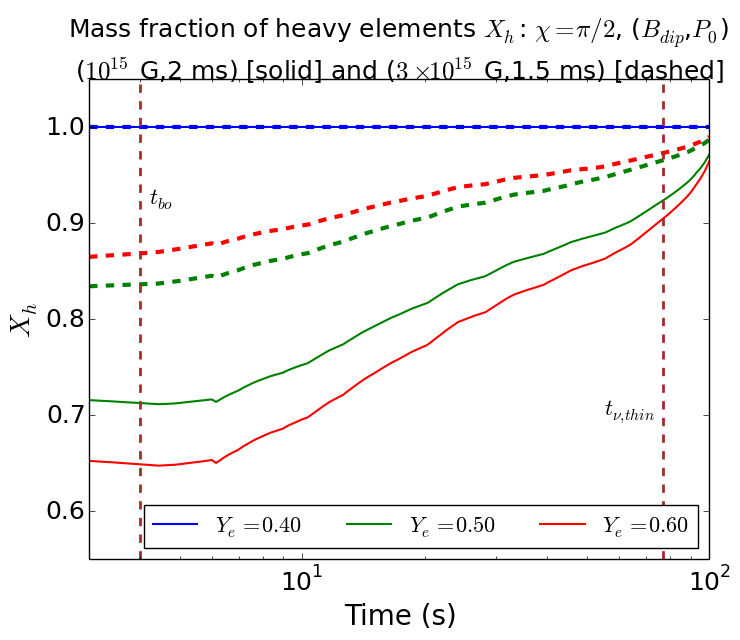}
\vspace{-0.5cm}
\caption{The time evolution of the mass fraction $X_h$ of heavy elements synthesised in PNS winds, shown for three electron fractions $Y_e=0.4$ (blue), $Y_e=0.5$ (green) and $Y_e=0.6$ (red). The results are calculated for $(B_{\rm dip},P_0,\chi)=(10^{15}\, {\rm G},2\, {\rm ms},\pi/2)$ [solid curves] and $(3\times10^{15}\, {\rm G},1.5\, {\rm ms},\pi/2)$ [dashed curves]. 
The breakout time corresponds to $(B_{\rm dip},P_0)=(10^{15}\, {\rm G},2\, {\rm ms})$.
} 
\label{Xh_vs_t}
\end{figure}

\begin{figure*} 
\includegraphics[width=0.49\textwidth]{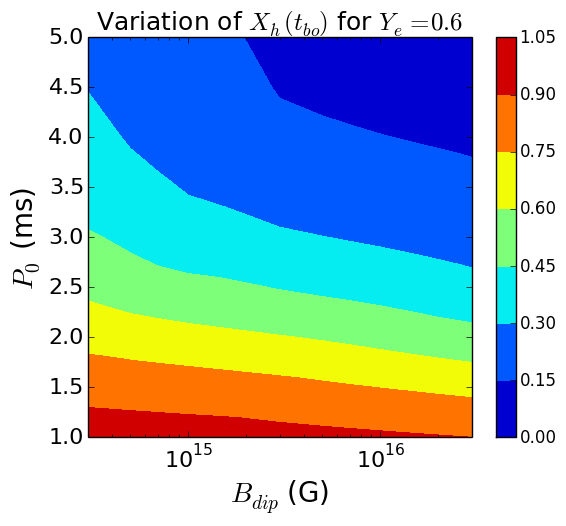}
\includegraphics[width=0.49\textwidth]{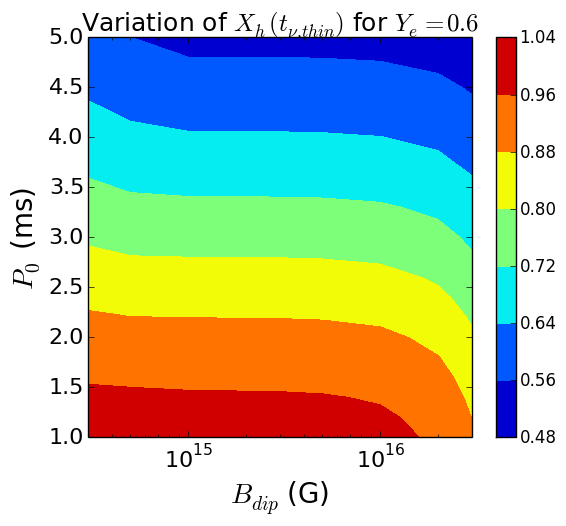}
\vspace{-0.3cm}
\caption{Contour plots for the mass fraction $X_h$ of heavy elements synthesised in PNS winds, shown as functions of surface dipole field $B_{\rm dip}$ and initial rotation period $P_0$. The left panel shows the results at the time of jet breakout ($t_{\rm bo}$) while the right panel shows the results at the neutrino transparency time ($t_{\nu,\rm thin}$), both for wind electron fraction $Y_e=0.6$ and fixed magnetic obliquity angle $\chi=\pi/2$.} 
\label{Ctr_Xh}
\end{figure*}

The electron fraction $Y_e$ is important primarily for two reasons. The value of $Y_e$ decides the channel through which He burns to form C. As this is the slowest reaction while forming the nuclei seeds, it essentially determines the neutron-to-seed ratio and therefore the total heavy element yield \citep{Hoffman1997,Pruet2006}. Moreover, $Y_e$ also determines the distribution of intermediate and heavy nuclei that is synthesised. In particular, for $Y_e \lesssim 0.5$, nuclei as heavy as $A \sim 90$ are formed via $\alpha$-particle and neutron capture reactions \citep{Roberts2010,AM2011}. If $Y_e$ is significantly lower, even heavier r-process elements can be potentially synthesised by additional neutron captures \citep{Seeger1965}. In contrast, under proton-rich conditions $Y_e \gtrsim 0.5$, primarily $A \sim 40-60$ elements are formed. As the specific value of $Y_e$ is sensitive to the uncertain details of neutrino emission and interactions \citep{RJ2000,Duan2011} and also depends on the PNS wind parameters, we consider a wide range $0.4 \lesssim Y_e \lesssim 0.6$ for our calculations. 

Given $S_{\rm wind}$, $t_{\rm exp}$ and $Y_e$, the total mass fraction $X_h$ that is synthesised in nuclei heavier than $A \gtrsim 56$ can be estimated as (see \citealt{Roberts2010})
\be 
X_h = \left\{
\begin{array}{ll}
1 - {\rm exp}\left[-8\times10^5\ Y_e^3 \left(\frac{t_{\rm exp}}{{\rm ms}}\right)\left(\frac{S_{\rm wind}}{k_b/{\rm nuc}}\right)^{-3}\right], & Y_e < 0.5 \vspace{0.2cm} \\
1 - \left[1 + 140(1-Y_e)^2 \left(\frac{t_{\rm exp}}{{\rm ms}}\right)\left(\frac{S_{\rm wind}}{k_b/{\rm nuc}}\right)^{-2}\right]^{-1/2}, & Y_e \geq 0.5 \\
\end{array}
\right.
\ee
In Figure \ref{Xh_vs_t}, the mass fraction in heavy nuclei $X_h$ is shown as a function of time following core bounce, for three values of the electron fraction $Y_e=0.4,\, 0.5,\, 0.6$. The results are shown for two wind models $(B_{\rm dip},P_0,\chi)=(10^{15}\, {\rm G},2\, {\rm ms},\pi/2)\, {\rm and}\, (3\times10^{15}\, {\rm G},1.5\, {\rm ms},\pi/2)$. It can be seen that for $Y_e \sim 0.4-0.6$, a significant fraction of the outflow mass is always locked into nuclei consisting of heavy elements. For a neutron-rich outflow with $Y_e=0.4$, $X_h \approx 1$ throughout the entire cooling phase as heavy elements can form very efficiently via the neutron capture reactions. However, the proton-rich outflows with $Y_e \sim 0.5-0.6$ tend to have a smaller heavy element mass fraction $X_h \lesssim 1$ even until late times as C must form through the slower triple-$\alpha$ reaction sequence. As larger field and rapid rotation lead to a smaller $S_{\rm wind}$ for similar $t_{\rm exp}$ (see Figure \ref{Swind_texp}), the heavy element mass fraction is slightly enhanced in such outflows. 

Figure \ref{Ctr_Xh} shows the contours for heavy nuclei mass fraction $X_h$ corresponding to $Y_e=0.6$, as functions of the surface dipole field $B_{\rm dip}$ and initial rotation period $P_0$. The left and right panel show $X_h$ computed at the jet breakout time and the neutrino transparency time, respectively. While the dependence of $X_h$ on $B_{\rm dip}$ is relatively weak, we find that $X_h$ increases substantially with rapid rotation. This is expected as a smaller $P_0$ will lead to smaller wind entropy for similar expansion timescales, thereby aiding heavy element nucleosynthesis. For very strong fields, $X_h$ is essentially suppressed due to the prominent magnetocentrifugal slinging effects. 
It can be seen that $X_h$ for a given $B_{\rm dip}$ and $P_0$ combination is larger at $t \approx t_{\nu,\rm thin}$ compared to $t \approx t_{\rm bo}$, which is expected as the mass fraction gradually rises over time due to an increasing $t_{\rm exp}$. We conclude that there is greater potential for the synthesis of heavier nuclei in outflows with rapid rotation rates and do not have excessively large magnetic fields. 

After the heavy nuclei are synthesised in the wind, they can still be destroyed at larger radii due to spallation which occurs when the nuclei collide with particles of relative energy greater than the nuclear binding energy $E_{\rm bind} \approx 8\, {\rm MeV/ nucleon}$. However, as the cross section for these inelastic collisions is similar to the Thomson cross section, they are only important for radii $r \ll r_{\rm ph} \sim 10^{11}\, {\rm cm}$ \citep{KG2007}. 
Furthermore, strong shocks that can potentially lead to the destruction of heavy nuclei are highly suppressed due to the large magnetisation of the PNS outflows \citep{KC1984}. Therefore, it is unlikely that the heavy nuclei that are synthesised in the wind will get destroyed during the jet collimation phase. The effect of spallation on the heavy nuclei survival (and thereby composition) during the jet collimation phase will be studied in more detail in a future work for the range of $B_{\rm dip}$ and $P_0$ considered here.

\subsection{Effect of PNS mass and obliquity angle on nuclei composition}
\label{Sec3.4}

\begin{figure*} 
\includegraphics[width=0.49\textwidth]{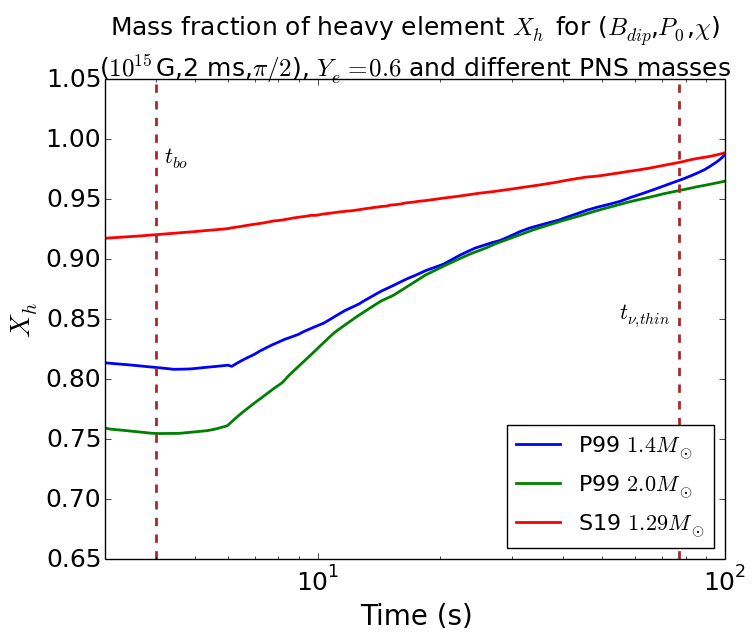}
\includegraphics[width=0.49\textwidth]{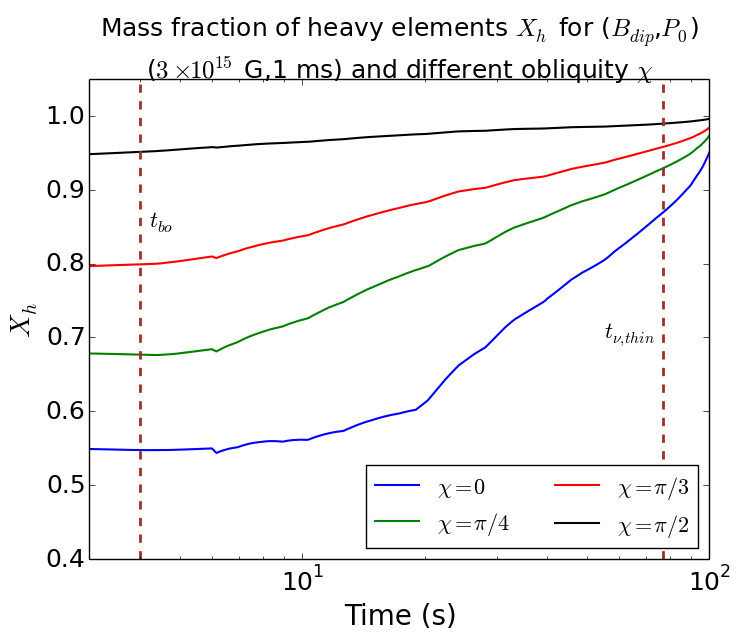}
\vspace{-0.3cm}
\caption{\emph{Left:} The effect of PNS mass $M_{\rm ns}$ on the evolution of heavy nuclei mass fraction $X_h$, shown for $(B_{\rm dip}, P_0, \chi) = (10^{15}\, {\rm G}, 2\, {\rm ms}, \pi/2)$, $Y_e=0.6$ and three PNS mass models: the low entropy $M_{\rm ns}=1.3\, M_{\odot}$ (M1L) models from \citet{Suwa2019} [labelled as S19] with the $M_{\rm ns}=1.4\, M_{\odot},\ 2.0\, M_{\odot}$ mass models from \citet{Pons1999} [labelled as P99]. 
\emph{Right:} The effect of magnetic obliquity angle $\chi$ on the time evolution of $X_h$, shown for $(B_{\rm dip}, P_0) = (3\times10^{15}\, {\rm G}, 1\, {\rm ms})$, $Y_e=0.6$ and four $\chi=0, \pi/4, \pi/3, \pi/2$. 
In both panels, the breakout time shown corresponds to the $(B_{\rm dip},P_0)=(10^{15}\, {\rm G},2\, {\rm ms})$ model.
}
\label{Xh_eff}
\end{figure*}

Here we discuss the effect of PNS mass and magnetic obliquity angle on the composition of UHECR nuclei. 
The left panel of Figure \ref{Xh_eff} shows the time evolution after core bounce of the heavy element mass fraction $X_h$ for wind model with $(B_{\rm dip},P_0,\chi)=(10^{15}\, {\rm G},2\, {\rm ms},\pi/2)$ and $Y_e=0.6$. For comparison, we use the low entropy model (M1L) from \citet{Suwa2019} with $M_{\rm ns}=1.29\, M_{\odot}$ and $M_{\rm ns}=1.4\, M_{\odot},\ 2.0\, M_{\odot}$ mass models from \citet{Pons1999}. 
We find that a higher PNS mass leads to smaller $\sigma_0$ and $t_{\rm exp}$ whereas the energy loss rate $\dot{E}_{\rm tot}$ increases marginally. This results in more photodisintegration of the synthesised nuclei at early times and thereby a smaller abundance of the synthesised heavy nuclei. The heavy nuclei mass fraction is fairly independent of the PNS mass at late times $\gtrsim 30\, {\rm s}$. 

The right panel of Figure \ref{Xh_eff} shows the evolution of heavy element mass fraction $X_h$ for wind model with $(B_{\rm dip},P_0)=(3\times10^{15}\, {\rm G},1\, {\rm ms})$, $Y_e=0.6$ and varying obliquity angle $\chi=0, \pi/3, \pi/4, \pi/2$. At fixed $B_{\rm dip}$ and $P_0$, the wind entropy $S_{\rm wind}$ tends to be more suppressed for an oblique rotator ($\chi=\pi/2$) as compared to the aligned rotator case. This can be explained due to the enhanced mass loss $\dot{M}$ that is caused by centrifugal slinging in the latter case. 
Furthermore, oblique outflows experience less neutrino heating, which directly suppresses the wind entropy.
As a result of this suppression in the wind entropy for relatively similar wind expansion timescales, the heavy element yield $X_h$ is enhanced significantly in more oblique rotators.

\section{Figure of merit parameter}
\label{Sec4}
In Section \ref{Sec3}, we estimated the mass fraction $X_h$ of heavy nuclei formed in outflows arising from strongly magnetised and rapidly rotating PNS. Here, we discuss the dynamical characteristics of the wind that are most important in determining the maximum mass for the synthesised nuclei. 
The maximum mass $A_{\rm max}$ to which the r-process nucleosynthesis can proceed depends on the ratio of free neutrons to seed nuclei following the completion of the $\alpha$-process.  
Therefore, the neutron-to-seed ratio depends sensitively on the entropy $S_{\rm wind}$, expansion timescale $t_{\rm exp}$ and electron fraction $Y_e$ at the radii where $\alpha$-particle formation concludes.  

Although the light r-process elements can be produced in abundance, recent studies of the neutrino-heated PNS winds \citep{Sumiyoshi2000,Thompson2001,Arcones2007,Roberts2010,Fischer2012} have shown that these outflows generally fail to reach the requisite neutron-to-seed ratio for the production of heavy r-process nuclei ($A \gtrsim 90$). The implication is that PNS outflows produce only the first r-process abundance peak and a different mechanism is responsible for the production of heavier (second and third peak) r-process elements (see e.g. \citealt{Eichler1989,Korobkin2012}). 

The fact that non-rotating and non-magnetic PNS winds fail to achieve the conditions required for the production of heavy nuclei is generally quantified in terms of the figure of merit parameter (see \citealt{Hoffman1997})
\be 
\xi_{\rm crit} = \frac{S_{\rm wind}^3}{Y_e^3 t_{\rm exp}} \approx 8\times10^9\ ({\rm k_b\ baryon^{-1})^3 s^{-1}}
\ee
For $\xi > \xi_{\rm crit}$, the nucleosynthesis can proceed to heavier elements but for $\xi < \xi_{\rm crit}$ it halts at smaller mass numbers \citep{MB1997}. 
It is important to note that $\xi_{\rm crit}$ can also vary depending on the underlying equation of state and/or more complicated $Y_e$ evolution. 
The typical entropy $50 \lesssim S_{\rm wind}/{\rm k_b\, baryon^{-1}} \lesssim 200$ and expansion timescale $0.01 \lesssim t_{\rm exp}/{\rm sec} \lesssim 0.1$ values for non-rotating and non-magnetic wind models indicate that $\xi = 2\times10^8 (S_{\rm wind}/100)^3 (t_{\rm exp}/0.01\, {\rm s})^{-1} (Y_e/0.5)^{-3}$ is significantly smaller than $\xi_{\rm crit}$. As the PNS gradually cools, both $S_{\rm wind}$ and $t_{\rm exp}$ increase, but the wind does not evolve into a state with $\xi > \xi_{\rm crit}$ \citep{QW1996,Thompson2001}. Models including additional heating sources \citep{QW1996,Metzger2007}, general relativity \citep{CF1997,Wanajo2001}, magnetocentrifugal acceleration \citep{Metzger2008} and supermassive NS \citep{Wanajo2013} have all been explored in an effort to bridge the gap between the $\xi$ values.

\begin{figure} 
\includegraphics[width=\columnwidth]{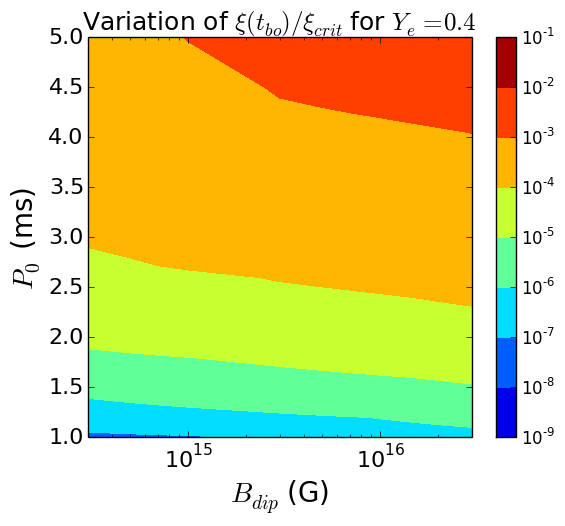}
\vspace{-0.6cm}
\caption{Contour plot for the figure of merit parameter $\xi$ relative to its critical value $\xi_{\rm crit} \approx 8\times10^9$, shown as a function of surface magnetic field $B_{\rm dip}$ and initial rotation period $P_0$, at the time of jet breakout $t_{\rm bo}$ and for electron fraction $Y_e=0.4$. } 
\label{xi_crit}
\end{figure}

Figure \ref{xi_crit} shows the $\xi/\xi_{\rm crit}$ contours evaluated at the jet breakout time $t_{\rm bo}$ as a function of the surface magnetic field $B_{\rm dip}$ and initial rotation period $P_0$, for a specific electron fraction $Y_e=0.4$. We find that the figure of merit parameter increases substantially for stronger dipole fields and slower PNS rotation rates, as can be seen from the top-right region of the contour plot. This is expected as the wind entropy $S_{\rm wind}$ is much less suppressed for slowly rotating PNS, although the expansion timescale $t_{\rm exp}$ turns out to be relatively similar. Even though $\xi/\xi_{\rm crit} \sim 10^{-2}$ can be obtained for the most favorable wind properties, these outflow conditions are still not sufficient to achieve the neutron-to-seed ratios required for the synthesis of heavy ($A \gtrsim 90$) r-process nuclei. Consequently, these outflows can only produce the first r-process abundance peak. \citet{Vlasov2014,Vlasov2017} found that PNS with strong $B \gtrsim 10^{14}\, {\rm G}$ and rapid rotation rate $P \lesssim {\rm few\, ms}$ produce outflows that are more favorable for the production of second and third peak r-process nuclei. This is primarily due to their significantly smaller expansion times through the seed nucleus forming region, although the entropy turns out to be moderately lower than the spherically non-rotating PNS winds. However, it should be noted that \citet{Vlasov2014,Vlasov2017} did not consider $S_{\rm wind}$ suppression due to magnetocentrifugal slinging effects and therefore obtained larger values for $\xi/\xi_{\rm crit}$, especially for rapidly rotating PNS with $P_0 \lesssim 2\, {\rm ms}$. Although \citet{Vlasov2014} showed that the critical ratio of $S_{\rm wind}^3/t_{\rm exp}$ was boosted up by a factor of $\sim 4-5$ compared to spherical outflows, it was still not sufficient for the nucleosynthesis to proceed to large $A \gtrsim 120$ based on the analytical criteria of \citet{Hoffman1997}.

\section{Nuclei acceleration and survival}
\label{Sec5}
During the first few seconds after core bounce, the PNS wind reaches mildly relativistic velocities ($\sigma_0 \lesssim 1$) since the neutrino-driven $\dot{M}$ is high. As the PNS gradually cools, the outflow becomes increasingly magnetically dominated and relativistic, with $\sigma_0 \gtrsim 10^2-10^3$ by $t \sim 20-50\, {\rm s}$. The conditions during this intermediate phase are ideal for both accelerating UHECRs and producing gamma-ray emission. The high-energy gamma-ray emission occurs when the jet dissipates its energy through internal shocks or magnetic reconnection at large radial distances $10^{13}\, {\rm cm} \lesssim r \lesssim 10^{16}\,{\rm cm}$. UHECRs are also likely to be accelerated by similar dissipation mechanisms that are responsible for powering GRB emission.  
At later times $t \gtrsim 100\ {\rm s}$, $\sigma_0$ increases even more rapidly as the neutrino-driven $\dot{M}$ drops abruptly once the PNS becomes transparent to neutrinos. Since jets with ultra-high magnetisation have difficulty in efficiently accelerating and dissipating their energy, this transition also ends the prompt GRB phase and hence UHECR production. In this section, we study the acceleration and survival of heavy nuclei that are synthesised in the PNS winds.

\begin{figure*} 
\includegraphics[width=0.55\textwidth]{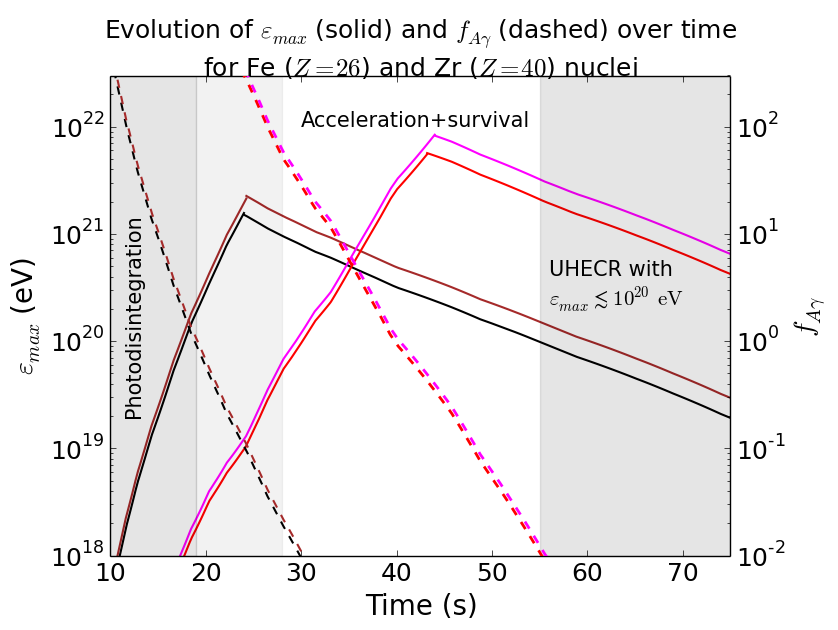}
\includegraphics[width=0.44\textwidth]{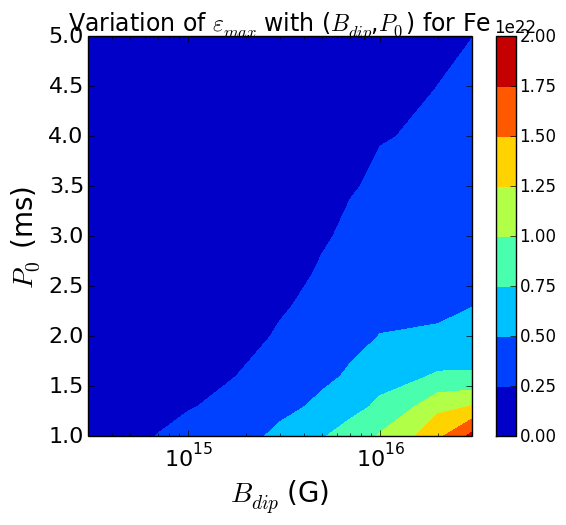}
\vspace{-0.3cm}
\caption{\emph{Left:} The maximum energy $\varepsilon_{\rm max}$ (obtained from $t_{\rm acc} \leq {\rm min}[t_{\rm cool},t_{\rm exp}]$, solid curves) to which the heavy nuclei can be accelerated and the effective photodisintegration optical depth $f_{A\gamma}$ (dashed curves), shown as functions of time for pure Fe ($Z=26$, black and red) and Zr ($Z=40$, brown and magenta) compositions. Two PNS parameter sets are shown: $(B_{\rm dip},P_0,\chi)=(10^{15}\, {\rm G},2\, {\rm ms},\pi/2)$ in black/brown, and $(3\times10^{15}\, {\rm G},1\, {\rm ms},\pi/2)$ in red/magenta. 
For $(10^{15}\, {\rm G},2\, {\rm ms})$, the unshaded area denotes the epoch during which the nuclei synthesised in the wind are simultaneously capable of reaching $\varepsilon_{\rm max}\gtrsim 10^{20}\, {\rm eV}$ and are yet not destroyed by the GRB photons ($0.02 \lesssim f_{A\gamma} \lesssim 1$ for $19 \lesssim t/{\rm sec} \lesssim 28$ and $f_{A\gamma} \lesssim 0.02$ for $t/{\rm sec} \gtrsim 28$). 
The corresponding epoch for $(3\times10^{15}\, {\rm G},1\, {\rm ms})$ starts at a later time interval and is not shown here.
\emph{Right:} The contour plot for the maximum energy $\varepsilon_{\rm max}$ to which Fe nuclei can be accelerated in the wind are shown as function of surface dipole magnetic field $B_{\rm dip}$ and initial rotation period $P_0$. Note that $\varepsilon_{\rm max}$ here is computed also ensuring the survival of nuclei.
} 
\label{Emax_vs_t}
\end{figure*}

We discuss the necessary conditions for the acceleration of UHECRs in magnetised outflows. For magnetically dominated PNS winds, most of the jet energy is contained in Poynting flux close to the central engine, which must be converted into bulk kinetic energy to explain the high Lorentz factors $\Gamma \gtrsim 10^2$ that are inferred from the GRB observations \citep{Waxman1995,LS2001,Bloom2003}.  Adopting the magnetic reconnection model developed by \citet{DS2002}, we assume that magnetic dissipation occurs gradually from a small outflow radius up to the saturation radius $r \sim R_{\rm mag} \approx 5\times10^{12}\, {\rm cm}\, (\sigma_0/10^{2})^{2} (P/{\rm ms}) (\epsilon/0.01)^{-1}$ beyond which the reconnection is complete. 
Here $v_r$ is the reconnection velocity and we assume a fixed $\epsilon = v_r/c \sim 0.01$ independent of radius or jet properties \citep{Lyubarsky2010,Uzdensky2010}. To constrain the maximum energy $\varepsilon_{\rm max}$ up to which cosmic rays can be accelerated, we consider first-order Fermi acceleration with a characteristic timescale similar to the Larmor gyration timescale $t_{\rm acc}=2\pi \eta_{\rm acc} E^{\prime}/ZeB^{\prime}c$, where $E^{\prime}=E/\Gamma$ and $B^{\prime} \approx (\dot{E}_{\rm iso}/r^2 \Gamma^2 c)^{1/2}$ are the particle energy and magnetic field strength in the jet frame, evaluated at $r=R_{\rm mag}$, $\Gamma \sim \sigma_0/2$ is the bulk Lorentz factor within the acceleration zone and $\eta_{\rm acc} \sim 1$ accounts for the uncertainty in the reconnection geometry. For the cosmic rays to be accelerated within the dynamical timescale of the outflow, $t_{\rm acc}$ must be smaller than the jet expansion timescale $t_{\rm exp} \approx R_{\rm mag}/\Gamma c$ \citep{Hillas1984}. 
The nuclei acceleration also competes with synchrotron cooling which proceeds on a timescale
$t_{\rm cool} = (0.4\, {\rm sec})\, \epsilon_{\rm mag}^{-1} (A/56)^4 (Z/26)^{-4} 
(\dot{E}_{\rm iso}/10^{52}\, {\rm erg\, s^{-1}})^{-1} (\Gamma/100)^3\\
(E/10^{20}\, {\rm eV})^{-1}
(r/10^{13}\, {\rm cm})^2$, where $\epsilon_{\rm mag} \sim 1$ is the fraction of jet power carried by Poynting flux. 

Heavy nuclei must avoid photodisintegration with the GRB photons in order to escape from the outflow. For isotropic photon fields, the photodisintegration interaction rate is given by (see, e.g., \citealt{Murase2008})
\be
\label{t_diss}
t_{A\gamma}^{-1}(\epsilon_A) = \frac{c}{2\gamma_{A}^2}\int_{\overline{\varepsilon}_{\rm th}}^{\infty}d\overline{\varepsilon}\ \overline{\varepsilon}\ \sigma_{A\gamma}(\overline{\varepsilon}) \int_{\overline{\varepsilon}/2\gamma_A}^{\infty}d\varepsilon \frac{1}{\varepsilon^2}\frac{dn}{d\varepsilon}
\ee
where $\sigma_{A\gamma}$ is the photodisintegration cross section, $\overline{\varepsilon}$ is the photon energy in the nucleus rest frame, $\overline{\varepsilon}_{\rm th}$ is the threshold energy and $\varepsilon_A=\gamma_A m_A c^2$ is the energy of the nuclei in the comoving frame. As the Band function characterises the spectrum of GRB prompt emission, the target photon spectrum $dn/d\epsilon$ can be effectively approximated by a broken power-law. For soft photons below the energy peak, the photodisintegration cross section can be approximated as $\sigma_{A\gamma}(\overline{\varepsilon}) \approx \sigma_{\rm GDR} \delta(\overline{\varepsilon} - \overline{\varepsilon}_{\rm GDR})\Delta \epsilon_{\rm GDR}$, where $\sigma_{\rm GDR}$, $\overline{\epsilon}_{\rm GDR}$ and $\Delta \epsilon_{\rm GDR}$ are the dipole resonance cross-section, energy and width, respectively. From Equation~\ref{t_diss}, the photodisintegration timescale simplifies to $t_{A\gamma}^{-1} \sim  (U_{\gamma}/5\varepsilon_p)c\sigma_{\rm GDR}(\Delta \epsilon_{\rm GDR}/\overline{\epsilon}_{\rm GDR})$, accounting for the bolometric correction \citep{KM2008a}. Here, $U_{\gamma}$ is the radiation energy density and $\varepsilon_p$ is the peak photon energy.

As the nuclei inelasticity is roughly $\kappa_{\rm GDR} \sim 1/A$ for the giant dipole resonance interactions, the resultant nuclei disintegration timescale is $t_{\rm diss}^{-1} = \kappa_{\rm GDR} t_{A\gamma}^{-1}$. The effective photodisintegration optical depth $f_{A\gamma}=t_{\rm exp}/t_{\rm diss}$, including the nuclei inelasticity factor, is then estimated as 
\be
f_{A\gamma} \approx \frac{\dot{E}_{\rm iso}\epsilon_{\rm rad}C\sigma_{\rm GDR} (\Delta \varepsilon_r/\overline{\varepsilon}_{\rm GDR})}{4\pi A \varepsilon_p rc\Gamma^2}
\ee
where $\epsilon_{\rm rad} \sim 0.1-1$  is the jet radiative efficiency, $C \sim 0.2$ is the fraction of the gamma-ray
photons that are released below the peak energy $\varepsilon_p \sim 0.1-1\, {\rm MeV}$. As giant dipole resonances dominate other energy loss mechanisms, we use a line width
$\Delta \epsilon_{\rm GDR}/\overline{\epsilon}_{\rm GDR} \sim 0.4 (A/56)^{0.21}$ and cross section $\sigma_{\rm GDR} \sim 8\times10^{-26}(A/56)\, {\rm cm^2}$ (see \citealt{Khan2005,KM2008a}). 
For nuclear survival, we consider both $f_{A\gamma} < 1$ (``effective'' survival) and $\tau_{A\gamma}<1$ ($f_{A\gamma} < 1/A$; ``complete'' survival). While the former implies that the nuclei do not lose all of their energy even after encountering dozens of interactions, the latter corresponds to the condition that the nuclei do not undergo any photodisintegration \citep{KM2010}. 
Considering $\tau_{A\gamma} \sim 1$ instead of $f_{A\gamma} \sim 1$, for photodisintegration to significantly affect the nuclei composition, naturally results in a marginally later epoch whereby the nuclei are also accelerated to higher energies.

The left panel of Figure \ref{Emax_vs_t} shows the maximum energy $\varepsilon_{\rm max}$ to which the heavy nuclei can be accelerated as a function of time, assuming energy dissipation via magnetic reconnection events in the outflow. While the black and red curves show the results for $(B_{\rm dip},P_0,\chi)=(10^{15}\, {\rm G},2\, {\rm ms},\pi/2)\, {\rm and}\, (3\times10^{15}\, {\rm G},1\, {\rm ms},\pi/2)$ for pure Fe composition, the brown and magenta curves show the corresponding results assuming pure Zr composition. The limiting energy $\varepsilon_{\rm max}$ is obtained based on the dynamical timescale constraint $t_{\rm acc} \leq t_{\rm exp}$ and the synchrotron cooling constraint $t_{\rm acc} \leq t_{\rm cool}$.
The dashed lines show the effective optical depth $f_{A\gamma}$ for the photodisintegration of heavy nuclei in the jet. It can be seen that $f_{A\gamma} \gg 10$ immediately after the jet breakout, as the heavy nuclei are disintegrated by the GRB photons. However, at later times, $f_{A\gamma} \propto \Gamma^{-2}r^{-1}$ decreases rapidly such that the nuclei can survive photodisintegration beyond $t \gtrsim 30\, {\rm s}$.

From the comparison of pure Fe ($Z=26$) and Zr ($Z=40$) composition results, we find that the time evolution of $\varepsilon_{\rm max}$ does not vary significantly with the wind composition. The unshaded region, shown for $(B_{\rm dip}, P_0, \chi)=(10^{15}\, {\rm G}, 2\, {\rm ms}, \pi/2)$, represents the epoch over which the Fe/Zr nuclei can achieve ultra-high energies $\varepsilon_{\rm max} \gtrsim 10^{20}\, {\rm eV}$ and yet not get disintegrated by the GRB photons i.e. $f_{A\gamma} \lesssim 1$. It can be seen that the onset of this epoch occurs at a slightly later time for stronger field and rapid rotation rate of the PNS central engine, by comparing the red/magenta curves with black/brown curves for pure Fe/Zr nuclear composition. 
Similarly, setting $\tau_{A\gamma} \sim 1$ instead of $f_{A\gamma} \sim 1$ results in a smaller time interval ($30 \lesssim t/{\rm sec} \lesssim 55$ as opposed to $20 \lesssim t/{\rm sec} \lesssim 55$) for the region where the nuclei can attain ultra-high energies and not get photodisintegrated.

The right panel of Figure \ref{Emax_vs_t} shows the contours for the maximum energy $\varepsilon_{\rm max}$ to which the Fe nuclei can be accelerated in the wind as function of both $B_{\rm dip}$ and $P_0$. We find that the limiting value of $\varepsilon_{\rm max}$ can be as large as $\sim10^{22}\, {\rm eV}$ for very strong fields $B_{\rm dip} \gtrsim 10^{16}\, {\rm G}$ and rapid rotation rates $P_0 \lesssim 1.5\, {\rm ms}$. This is in agreement with the result shown in the left panel as the peak of the $\varepsilon_{\rm max}$ curve shifts to larger energy (and time) for the strong field and rapid rotation model. 
The limiting energies for all combinations of $B_{\rm dip}$ and $P_0$ are evaluated after ensuring the survival of nuclei in the outflow i.e. with photodisintegration optical depth $f_{A\gamma}<1$.
Therefore, the central engines with large $B_{\rm dip}$ and small $P_0$ are more likely to explain the observed high energies of UHECRs.

\section{Discussion \& Implications}
\label{Sec6}
As the PNS gradually cools over Kelvin-Helmholtz timescale $t_{\rm KH} \sim {\rm a~few}\, 10~{\rm s}-100\, {\rm s}$, changes in the wind properties are largely determined by the increase in the magnetisation $\sigma_0$. 
While higher $P_0$ increases the $\sigma_0$ growth rate, $B_{\rm dip}$ only affects the initial value of the wind magnetisation. 
For a rapidly spinning PNS, $\sigma_0$ is strongly suppressed due to the centrifugally accelerated wind mass-loss rate. 
The presence of strong surface fields and rapid rotation rates both facilitate an increase in the wind energy loss rate $\dot{E}_{\rm tot}$, which in turn increases the feasibility of successful GRB jets. 

In order for nucleosynthesis to occur in the wind, the nuclei that are initially loaded must be disintegrated into free nucleons. 
Our analysis validates that the photodisintegration and spallation interactions are indeed significant for the initially loaded nuclei, such that they get rapidly disintegrated into free nucleons which are then utilised as the seeds for subsequent nucleosynthesis.  
Compared to GRB fireball jets, the strong fields in magnetically-powered outflows suppress the wind entropy considerably, which enables intermediate and heavy nuclei to form efficiently and results in higher asymptotic elemental yields. 
We estimated $X_h \approx 1$ for neutron-rich outflows ($Y_e \approx 0.4$) throughout the cooling phase, which is due to efficient synthesis via neutron-capture reactions.
Moderately proton-rich outflows ($Y_e \approx 0.6$) still exhibit $X_h \gtrsim 0.5$ even until late times as C forms through the slower triple-alpha reaction sequence. Overall, we find that magnetised outflows generated from PNS with smaller fields and rapid rotation provide greater potential for the synthesis of heavy nuclei. 

The maximum mass $A_{\rm max}$ to which nucleosynthesis can proceed is strongly determined by the free neutron to seed nuclei ratio, which depends on the outflow parameters $S_{\rm wind}$, $t_{\rm exp}$ and $Y_e$, and is quantified by the figure of merit parameter $\xi$.  
We find that the outflows with large $B_{\rm dip}$ as well as $P_0$ provide favourable conditions but still cannot achieve the typical neutron to seed nuclei ratios necessary for the synthesis of heavy nuclei $A \gtrsim 90$. The primary implication is that these magnetised outflows can only produce the first r-process abundance peak while a different mechanism is responsible for the production of heavier r-process elements. Recent works \citep{Vlasov2014,Vlasov2017} have shown that outflows traversing along the intermediate magnetic field lines for an aligned rotator ($\chi=0$) present a promising scenario for r-process nucleosynthesis. Considering a broad range of magnetic obliquity angle $0 \lesssim \chi \lesssim \pi/2$, we have shown that the oblique rotators ($\chi=\pi/2$) are generally more conducive for heavy nucleosynthesis due to their suppressed wind entropy. However, even with these optimistic geometries, it is difficult to generate nuclei that are heavier than the second r-process peak (see Figure \ref{xi_crit}). 
Detailed abundance patterns for magnetised outflows generated from PNS systems are beyond the scope of this paper, but can be explored by nuclear reaction network calculations such as SkyNet \citep{LR2017}.

UHECRs are likely to be accelerated by a similar dissipation mechanism which is also responsible for powering GRB emission, provided that magnetar central engines also power GRBs. 
Considering magnetic reconnection models for energy dissipation powering GRB emission 
\citep{DS2002}, we have shown that during the cooling epoch $25 \lesssim t/{\rm sec} \lesssim 55$, the synthesised heavy nuclei can simultaneously achieve ultra-high energies $\varepsilon_{\rm max} \gtrsim 10^{20}\, {\rm eV}$ and still not get disintegrated by GRB photons. The onset of this epoch occurs at marginally later times for stronger fields and rapid rotation rates. 
We find that the evolution of $\varepsilon_{\rm max}$ does not vary appreciably with a change in wind composition. Furthermore, the limiting value of $\varepsilon_{\rm max}$ that is obtained ensuring nuclei survival, can be as high as $\sim 10^{22}\, {\rm eV}$ for sufficiently strong fields $B_{\rm dip} \gtrsim 10^{16}\, {\rm G}$ and rapid rotation $P_0 \lesssim 1.5\, {\rm ms}$, which justifies the observed energy range of UHECRs. 

Lastly, we have made some physical assumptions in order to simplify our modelling of the magnetar cooling evolution and wind nucleosynthesis. Although we adopt a fixed stretch factor $\eta_s \approx 3$ to modify the rotating PNS cooling evolution relative to non-rotating case, this factor should ideally be an increasing function of the rotation rate. 
The NS EoS and the magnetic obliquity $\chi$ can both affect the nucleosynthesis yields that are estimated from our calculations (see Section \ref{Sec3.4} for related discussion). 
While late-time accretion onto the PNS can affect its spin down evolution by intensifying the neutrino-driven mass loss via centrifugal slinging, $B_{\rm dip}$ may also be dynamically amplified over $t \sim t_{\rm KH}$ due to differential rotation (see Appendix \ref{AppendixB}). 
We have also assumed a pure ejecta composition and electron fraction $Y_e \sim 0.4-0.6$ whereas they can be sensitive to the neutrino interactions occurring within the outflow and may vary based on the central engine properties.  
Finally, internal shocks resulting from the collision of mass shells can also be a plausible mechanism for jet energy dissipation and UHECR acceleration.

\section{Summary \& Conclusions}
\label{Sec7}
The composition of UHECRs provides crucial information regarding the origin of these energetic particles. Recent detections of the UHECR spectra made by Pierre Auger Observatory suggest that the cosmic-ray composition is dominated by intermediate/heavy nuclei at high energies. 
In this paper, we have presented a comprehensive study of the steady-state properties of neutrino-driven winds originating from rapidly rotating proto-magnetars shortly after core collapse. We examined in detail the synthesis, acceleration and survival of UHECR nuclei in outflows that emerge from the strongly magnetised and rapidly rotating PNS. For our analysis, we have considered a broad range of central engine properties, namely central engines with mass $1.3\, M_{\odot} \lesssim M_{\rm ns} \lesssim 2.0\, M_{\odot}$, dipole magnetic fields $3\times10^{14}\, {\rm G} \lesssim B_{\rm dip} \lesssim 3\times10^{16}\, {\rm G}$, rotation periods $1\, {\rm ms} \lesssim P_0 \lesssim 5\, {\rm ms}$ and obliquity angle $0 \lesssim \chi \lesssim \pi/2$. 

As UHECR sources still remain uncertain, their composition provides a useful probe towards constraining the potential candidates. If the UHECR nuclei composition at high energies is dominated by intermediate/heavy nuclei, as indicated by the recent PAO measurements, the relativistic winds generated from millisecond magnetars can be promising sources as their environments consist of large fractions of intermediate/heavy nuclei. Besides their role as plausible nucleosynthesis sites, magnetically-dominated outflows from millisecond proto-magnetars are also prime contenders for GRB central engines. 
Despite the simplifications in our analytical model, we were able to show that UHECR nuclei can be synthesised and accelerated within the jet to $\varepsilon_{\rm max} \gtrsim 10^{21}\, {\rm eV}$ over a broad range of $B_{\rm dip}$ and $P_0$, making neutrino-driven winds from protomagnetar central engines promising sources of UHECR.

Although their energies range over many orders of magnitude, the comparable energy fluxes of high-energy neutrinos, UHECRs and gamma-rays suggests a physical link among these three messengers. In particular, \citet{Murase2008} have shown that the conditions for heavy UHECR nuclei to survive is the same as those for GeV-TeV gamma rays to emanate and for TeV-PeV neutrinos to be not produced. Therefore, future multi-messenger studies will further help resolve the sources of UHECRs.

\section*{Acknowledgements}
We thank Kunihito Ioka, Brian Metzger and B.~Theodore Zhang for comments on the manuscript, and Nick Ekanger for discussions. We would like to thank Yudai Suwa for generously providing the simulation results of the low-entropy $M_{\rm ns}=1.3\, M_{\odot}$ models from \citet{Suwa2019}. M.B.~is supported by NSF Research Grant No.~AST-1908960, and by Eberly Research Fellowship at the Pennsylvania State University. S.H.~is supported by NSF Grant No.~AST-1908960 and No.~PHY-1914409, and by the U.S. Department of Energy Office of Science under award number DE-SC0020262. This work was supported by World Premier International Research Center Initiative (WPI), MEXT, Japan. The work of K.M. is supported by the NSF Grant No.~AST-1908689, No.~AST-2108466 and No.~AST-2108467, and KAKENHI No.~20H01901 and No.~20H05852.

\section*{Data availability}
The data underlying this article will be shared on reasonable request to the corresponding author.

\bibliography{main}

\appendix 

\section{Neutrino cooling models}
\label{AppendixA}
For our calculations, we use $L_{\nu}(t)$, $\epsilon_{\nu}(t)$ and $R_{\rm ns}(t)$ from \citet{Pons1999} [hereafter P99] who computed the cooling evolution for non-rotating PNS \citep{BL1986}. 
During the initial cooling phase $t \sim {\rm few}\ {\rm s}$, $L_{\nu}$ and $\epsilon_{\nu}$ gradually decrease with a power-law dependence while the PNS also shrinks in size from its initial radius $\sim 30\, {\rm km}$ to the final radius $\sim 10\, {\rm km}$. 
As the degenerate matter cools further, the PNS becomes transparent to the neutrino emission around $t = t_{\nu,\rm thin} \sim 77\ {\rm s}$ after which both $L_{\nu}$ and $\epsilon_{\nu}$ drop sharply. During this transition from diffusion dominated phase to the optically thin cooling phase, $L_{\nu}$ drops by over an order of magnitude as the neutrinosphere retracts into the neutron star. 

The effects of magnetic field, rotation and convective transport are not included in these long-term cooling calculations (see Section \ref{Sec2.2}). 
Furthermore, the precise value of $t_{\nu,\rm thin}$ also depends on the uncertain details of the high density neutron star EoS. 
To consider the rotational effects qualitatively, we follow \cite{Metzger2011a} and adopt a stretch parameter $\eta_s \sim 3$ that modifies the cooling evolution but also preserves the total energy content in the neutrinos (see equation 3).

\begin{figure} 
\includegraphics[width=\columnwidth]{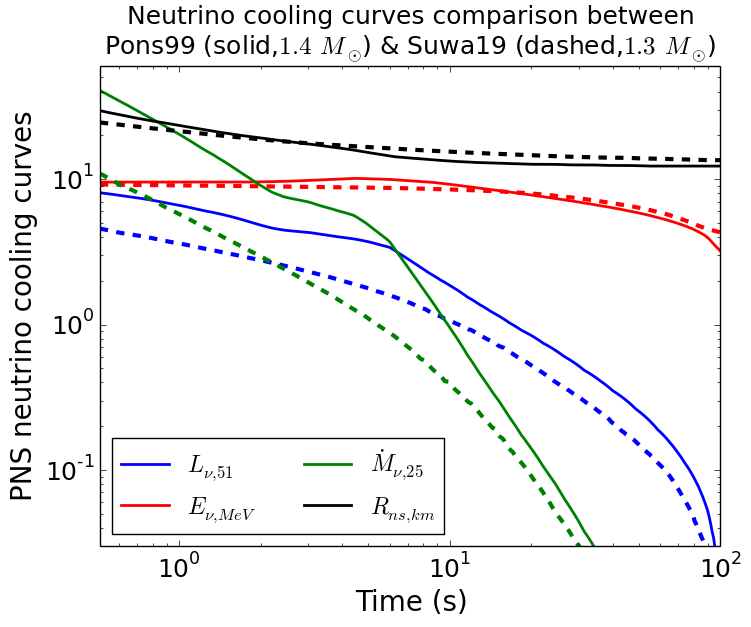}
\vspace{-0.5cm}
\caption{The time evolution of neutrino luminosity (blue) and energy (red), PNS radius (black) and the resultant wind mass loss rate (green) are shown for a $M_{\rm ns}=1.4\, M_{\odot}$ PNS, as obtained from \citet{Pons1999}  simulation results (solid curves). Here $L_{\nu}$ is shown in units of $10^{51}\, {\rm erg/s}$, $E_{\nu}$ in MeV, $R_{\rm ns}$ in km and $\dot{M}_{\nu}$ is obtained from equation (\ref{Mdot}) normalised to $10^{25}\, {\rm CGS\, units}$.  
For comparison, we also show the low entropy $M_{\rm ns}=1.3\, M_{\odot}$ results (dashed curves) from \citet{Suwa2019} -- see their Figure \ref{Swind_texp}.
}
\label{Fig_neuLE}
\vspace{-0.2cm}
\end{figure}

In order to explore the sensitivity of the results to uncertainties in $L_{\nu}$, $\epsilon_{\nu}$ and $R_{\rm ns}$ evolution, we also use neutrino cooling curves from the recent calculations of \citet{Suwa2019} [hereafter S19]. In Figure \ref{Fig_neuLE}, we compare the P99 cooling curves with the low entropy $M_{\rm ns}=1.3\, M_{\odot}$ (M1L) models of S19. While the $R_{\rm ns}$ and $\epsilon_{\nu}$ evolutions are fairly similar over the entire cooling phase, the primary difference between the neutrino cooling curves is that $L_{\nu}$ and therefore $\dot{M}_{\nu}$ is suppressed at early times $t \lesssim 20\, {\rm s}$ for the S19 result. 
As the residual values of $S_{\rm wind}$ and $t_{\rm exp}$ are roughly comparable, the heavy nuclei composition $X_h$ is not found to be much different compared to those from the P99 simulations. 

\section{Effect of accretion and differential rotation}
\label{AppendixB}
Shortly after core collapse, a magnetically powered outflow develops outwards from the PNS. At early times, the explosion asymmetry allows for continued matter accretion through an equatorial disk \citep{Vartanyan2019,Burrows2020}. Although the accretion rate onto the PNS drops substantially at later times, it does not go down to zero. A fraction of the gravitational binding energy released from the fallback of this material is converted into neutrino emission. This accretion induced emission competes with the diffusion luminosity from the PNS. The NS remains rapidly spinning by accreting angular momentum, which enhances mass loss from higher latitudes due to magnetocentrifugal slinging \citep{Thompson2004,Metzger2007}. Here we consider how the mass accretion at late times onto the PNS can affect its spin-down evolution \citep{Metzger2008,ZD2009}.

The rapidly rotating PNS ejects a considerable fraction of its mass into a disc during the core collapse in order to conserve angular momentum. \citet{Dessart2006} performed 2D magnetohydrodynamic simulations to show that a quasi-Keplerian accretion disc of mass $\sim 0.1-0.5\, M_{\odot}$ forms around the PNS, extending from the surface at $R_{\rm ns} \sim 30\, {\rm km}$ to large radii. The characteristic timescale for accretion to occur is the viscous timescale $t_{\rm visc} = R^2/\alpha\Omega_K H^2 = 1\, {\rm s}\,
(M/M_{\odot})^{-1/2} (\alpha/0.1)^{-1}(R_0/4R_{\rm ns})^{3/2}(H/0.2R_0)^{-2}$. Here $\alpha$ is the viscosity parameter, $M$ is the PNS mass; $R_0$, $H$ and $\Omega_K = (GM/R_0^3)^{1/2}$ are the disc's radius, scaleheight and Keplerian rotation rate, respectively. On somewhat longer timescales, the disk mass and accretion rate gradually decrease, and the PNS spins up through accretion as well as by its KH contraction. 

If the magnetic field is amplified on a timescale comparable to the duration of the NS cooling epoch, the assumption of a fixed dipole flux may not be a good approximation. Differential rotation can be a source of energy to power field growth via $\alpha-\Omega$ dynamo in the convective PNS or magnetorotational instability. Magnetic field generated from the energy available in differential rotation can be written as (see \citealt{Metzger2011a})
\be
B_{\rm dip} = 10^{16}\ {\rm G} \left(\frac{\epsilon_B}{10^{-3}}\right)^{1/2} \left(\frac{R_{\rm ns}}{12\ {\rm km}}\right)^{-1/2}\left(\frac{P_0}{{\rm ms}}\right)^{-1}
\ee
assuming that magnetic energy in the dipole field is a fraction $\epsilon_B$ of the rotational energy. 

\begin{figure} 
\includegraphics[width=\columnwidth]{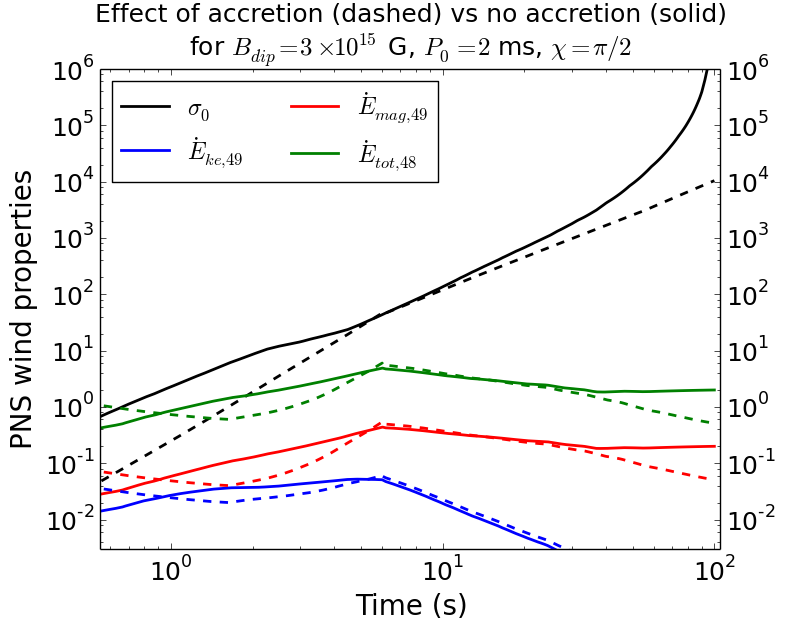}
\vspace{-0.5cm}
\caption{The effect of mass accretion from the disk and magnetic field growth due to PNS differential rotation, on the wind magnetisation $\sigma_0$ and energy loss rate $\dot{E}_{\rm tot}$, is shown for $(B_{\rm dip},P_0,\chi) = (3\times10^{15}\, {\rm G},2\, {\rm ms},\pi/2)$. The results including accretion and differential rotation are shown as dashed curves.
} 
\label{Fig_massacc}
\vspace{-0.2cm}
\end{figure}

The rate at which the disk's angular momentum is accreted by the PNS is $\dot{J}_D = \dot{M}_D R_0^2 \Omega_K$, where $\dot{M}_D$ is the mass accretion rate, $R_0 = 14 R_g \approx 60\, {\rm km}$ is a fixed cylindrical radius and $\Omega_K = (GM_{\rm ns}/R_0^3)^{1/2}$ is the Keplerian rate at which the outflow base rotates \citep{Metzger2008}. During the contraction phase of the PNS over the first few seconds, the total angular momentum is roughly conserved
\be
J_{\rm tot} = J_{\rm PNS} + J_{\rm disk} = (2/5)M_{\rm ns}R_{\rm ns}^2 \Omega + \eta t_{\rm visc} \dot{M}_D R_0^2 \Omega_K 
\ee
where $\eta \approx 0.1$ denotes the fraction of the mass that accretes angular momentum onto the PNS and $\dot{M}_D t_{\rm visc} \sim 0.1 M_{\rm ns}$ is the accreted mass assuming $\dot{M}_D \sim 0.1\, M_{\odot}/s$ and $t_{\rm visc} \sim 1\ {\rm s}$. 
The spin-down evolution of the PNS and disk system is then obtained from $\dot{J}_{\rm tot} = -\dot{E}_{\rm tot}/\Omega$, where $\dot{E}_{\rm tot} = (\dot{M}_{\rm PNS} + \dot{M}_{\rm disk})c^2 \sigma_{\rm tot}$ depends on total magnetisation parameter
\be
\sigma_{\rm tot} = \frac{(\phi_{B,\rm PNS} + \phi_{B,\rm disk})^2}{c^3(\dot{M}_{\rm PNS}/\Omega_{\rm PNS}^2 + \dot{M}_{\rm disk}/\Omega_{\rm disk}^2)}
\ee
where $\phi_{B,\rm PNS/disk}$ is the magnetic flux, $\Omega_{\rm PNS/disk}$ is the rotation rate and $\dot{M}_{\rm PNS/disk}$ is the mass loss rate for the PNS/disk respectively. To compute $\dot{M}_{\rm disk}$, we estimate the early-time disk neutrino luminosity $L_{\nu,\rm disk}=10^{52}(t/1\, {\rm sec})^{-1}\, {\rm erg/s}$ and energy $\epsilon_{\nu,\rm disk}=10\, {\rm MeV}$ in agreement with the analytical results obtained by \citet{Metzger2008}. 

Figure \ref{Fig_massacc} shows the effect of mass  accretion from an equatorial disk as well as $B_{\rm dip}$ growth aided by differential rotation on the long-term PNS cooling evolution, particularly $\sigma_0$ and $\dot{E}_{\rm tot}$ of the outflow. In the presence of accretion and differential rotation, we find that the wind magnetisation is suppressed at early times $t \lesssim {\rm few\, s}$ after the core bounce and $\sigma_0$ exhibits a slower growth rate for late times $t \sim {\rm few}\, 10{\rm s}$. This is expected as the contribution from the disk mass loss term $\dot{M}_{\rm disk}/\Omega_{\rm disk}^2$ to the denominator more than compensates for the additional magnetic flux $\phi_{B,\rm disk}$ from the equatorial disk.  Similarly, as $\dot{M}_{\rm disk} \ll \dot{M}_{\rm PNS}$, the wind energy loss rate $\dot{E}_{\rm tot} \propto \dot{M}_{\rm PNS}\sigma_{\rm tot}$ also turns out to be smaller at early times (primarily due to the smaller effective $\sigma_{\rm tot}$) with a faster fall off rate at late times. As the wind $\sigma_0$ and $\dot{E}_{\rm tot}$ are essentially similar during the epoch when most of the nucleosynthesis occurs in the wind, the heavy element yields remain unaffected by these effects.  

\label{lastpage}

\end{document}